\documentclass[useAASTex,usenatbib,fleqn]{mn2e}

\usepackage[english]{babel}
\usepackage{graphicx}
\usepackage{fleqn}
\usepackage{amsfonts}
\usepackage{amssymb}
\usepackage{amsmath}
\usepackage{booktabs}
\usepackage{hyperref}
\usepackage{times}
\date{Accepted 2014 October 8.  Received 2014 September 23; in original form 2014 June 26}

\newcommand{\Ec}{\mathcal{E}} 
\newcommand{\Rc}{\mathcal{R}} 
\title{Probing the formation of planetesimals in the Galactic Centre using Sgr A* flares}
\author[Adrian S. Hamers and Simon F. Portegies Zwart]{Adrian S. Hamers$^{1}$ and Simon F. Portegies Zwart$^{1}$ \\
$^{1}$Leiden Observatory, Leiden University, PO Box 9513, NL-2300 RA Leiden, The Netherlands \\}

\begin{document}

\maketitle

\begin{abstract}
Flares in X-ray and near infrared are observed above the quiescent emission of the supermassive black hole (SBH) in the Galactic Centre (GC) at a rate of approximately once per day. One proposed energy source for these flares is the tidal disruption of planetesimals with radius $\gtrsim 10$ km passing within $\sim 1 \, \mathrm{AU}$ of the SBH. Very little is known about the formation and evolution of planetesimals in galactic nuclei such as the GC, making predictions for flaring event rates uncertain. We explore two scenarios for the formation of planetesimals in the GC: (1) in a large-scale cloud bound to the SBH, and (2) in debris discs around stars. We model their orbital evolution around the SBH using the Fokker-Planck equation and investigate the effect of gravitational interactions with various relevant perturbers. Our predicted flaring rate, $\approx 0.6 \, \mathrm{day^{-1}}$, is nearly independent of the distribution of perturbers. Moreover, it is insensitive to scenarios (1) or (2). The assumed number of planetesimals per star is consistent with debris discs around stars in the Solar neighbourhood. In scenario (1) this implies that the number of planetesimals formed in the large-scale cloud is strongly correlated with the number of stars, and this requires finetuning for our results to be consistent with the observed flaring rate. We favour the alternative explanation that planetesimals in the GC are formed in debris discs around stars, similar to the Solar neighbourhood.  
\end{abstract}

\section{Introduction}
\label{sect:introduction}
The Galactic Centre (GC) contains a supermassive black hole (SBH) of mass $M_\bullet \approx 4 \times 10^6 \, \mathrm{M_\odot}$ \citep{ghez_ea08,gillessen_ea09}. For the last decade, observations in the near infrared and X-ray of the central region of the GC have revealed the existence of flares, occurring approximately once per day \citep{baganoff_ea01,baganoff_ea03,genzel_ea03,dodds-eden_ea11,barriere_ea14}. These flares are 3-100 times more luminous than the quiescent emission of the central radio source Sgr A*, which is known to be very dim: its bolometric luminosity is only $\sim 10^{-8.5} L_\mathrm{Edd}$ \citep{geg10}. Observations indicate that the region from which the flares originate is very compact, extending no more than a few tens of gravitational radii $r_g=GM_\bullet/c^2$ from the SBH \citep{baganoff_ea01,genzel_ea03,porquet_ea03,shen_ea05}. 

Several explanations for the energy source and emission mechanism of the flares have been proposed. One of these is the tidal disruption of planetesimals with radius $\gtrsim 10 \, \mathrm{km}$ (\citealt{znm12}, hereafter ZNM12). ZNM12 showed that if the latter passes within $\sim 1 \, \mathrm{AU}$ of the SBH, it is broken up into smaller fragments by tidal forces; the fragments subsequently vaporise because of friction with the ambient gas. When the vaporised material is mixed with the accretion flow on to Sgr A*, enough energy could be released to produce an observable flare with X-ray luminosity $\sim 10^{34}-10^{35} \, \mathrm{erg/s}$. In ZNM it was assumed that the planetesimals were formed in debris discs around stars and were subsequently stripped by the tidal force of the SBH and by gravitational encounters with other stars. The resulting event rate of the flares was estimated using loss cone refilling arguments and a rate was found that is consistent with the observed rate of approximately once per day. The method employed by ZNM12 to calculate the flaring rate was very approximate; details about the distribution of stars in the GC and the spatial and temporal dependence of the stripping of planetesimals were not taken into account. 

Very little is known about the formation and evolution of such planetesimals in galactic nuclei like the GC. One possibility is that they are formed in a large-scale spherical cloud orbiting the SBH. Another possibility is that they are born in debris discs around stars (see e.g. \citealt{wyatt08} for a review), and are stripped by the tidal force of the SBH or gravitational encounters with other stars (e.g. \citealt{nss12}). In this paper we investigate both scenarios by means of numerical integrations of the Fokker-Planck equation. We model the orbital energy evolution around the SBH, taking into account the effects of gravitational perturbations from late-type stars, a possible cusp of stellar black holes close to the SBH and giant molecular clouds further away, as well as the effects of physical collisions. 

We will show that the predicted present-day disruption rates in the GC differ very little between the two scenarios and that this conclusion depends weakly on the details of the perturbers. In both scenarios, we find a disruption rate of $\sim 1 \, \mathrm{day^{-1}}$ assuming that the number of planetesimals per (late-type) star is $N_\mathrm{a/\star}=2\times 10^7$. The number $N_\mathrm{a/\star}=2\times 10^7$ is consistent with debris discs observed around stars in the Solar neighbourhood. In the first scenario, in which the planetesimals are formed in a large cloud, this implies that the number of bodies formed is strongly correlated with the number of stars, and this requires finetuning of the quantity $N_\mathrm{a/\star}$. We favour the more natural explanation that planetesimals in galactic nuclei similar to the GC are formed in debris discs around stars, no differently than planetesimals around stars in the Solar neighbourhood. 

The structure of this paper is as follows. In Section~\ref{sect:models} we describe our models of the GC and of debris discs around stars, in Section~\ref{sect:stripping} we describe the stripping process and in Section~\ref{sect:dyn_ff_ast} we present our integrations of the Fokker-Planck equation and the implied disruption rates. We extrapolate our results to different galactic nuclei in Section~\ref{sect:discussion} and briefly discuss a special case that would lead to a burst of flares. We conclude in Section~\ref{sect:conclusions}.

\section{Setting the stage}
\label{sect:models}
Before describing the effects of the stripping of planetesimals from stars (Section~\ref{sect:stripping}) and presenting our main results of their orbital evolution around the SBH (Section~\ref{sect:dyn_ff_ast}) we discuss our models of the GC (Section~\ref{sect:models:GC}) and of the adopted initial debris discs around stars (Section~\ref{sect:models:disc_models}).

\subsection{Models of the GC}
\label{sect:models:GC}
Here, we describe and motivate the models of the GC on which the calculations in Section~\ref{sect:dyn_ff_ast} are based. For further details on the implementation of these models in the integrations of the Fokker-Planck equation (Section~\ref{sect:dyn_ff_ast}) we refer to Appendix \ref{app:FP_main}.

\subsubsection{Late-type stars}
\label{sect:models:GC:LT}
The majority (96\%) of the observed stars in the GC are old ($> \, \mathrm{Gyr}$) late-type stars \citep{geg10}. Number counts indicate that the 3D number density $n(r)$ of these stars approximately scales with the distance to the SBH as $n(r) \propto r^{-1.8}$ for $r\gtrsim 0.3 \, \mathrm{pc}$, whereas there is a much flatter, possibly even declining density profile inside $\sim 0.3 \, \mathrm{pc}$ \citep{eckart_ea93,genzel_ea96,schodel_ea07,trippe_ea08,bse09,schodel_ea09,ohf09,do_ea13}. There is evidence that at large radii the density drops more steeply than $n(r) \propto r^{-1.8}$, i.e. $n(r) \propto r^{-3}$ for $r\gtrsim 5 \, \mathrm{pc}$ \citep{schodel_ea14}. 

Most of the late-type stars likely formed $\sim 10 \, \mathrm{Gyr}$ ago, possibly coincident with the Galactic Bulge \citep{blum_ea03,maness_ea07,pfuhl_ea11}. From theoretical arguments these stars are expected to be distributed in a cusp, with number density $n(r) \propto r^{-7/4}$, after approximately a relaxation time-scale \citep{bw76}. This is consistent with the observed distribution of late-type stars at radii $0.3\,\mathrm{pc} \lesssim r \lesssim 5 \, \mathrm{pc}$. At smaller radii, however, a core is observed. The nature of this core is not fully understood, but a possible explanation is that the relaxation time-scale in the GC at the radius of influence is likely $> 10 \, \mathrm{Gyr}$, therefore the core could  reflect the primordial population of the late-type stars \citep{merritt10}. \citet{merritt10} showed that if there initially was a core of late-type stars of size $\sim 1\, \mathrm{pc}$, this would have evolved to the observed size of $\sim 0.3 \, \mathrm{pc}$ today. 

It is conceivable that the observed core of late-type stars arises from observational bias: within $0.3 \, \mathrm{pc}$ the stellar light is dominated by bright early-type stars (see below), which complicates observations of late-type stars. To accommodate both possibilities (i.e. a core versus a cusp) we consider the following two (spherically symmetric) models of the present-day number density $n_\mathrm{LT}(r)$ of late-type stars in the GC (`LT1' and `LT2'),
\begin{align}
&\nonumber n_\mathrm{LT}(r) = \\
& \left \{ \begin{array}{ll}
n_b \left ( \frac{r}{r_b} \right )^{-\gamma_i} \left[ 1 + \left( \frac{r}{r_b} \right )^\alpha \right]^{\frac{\gamma_i-\gamma}{\alpha}}\left(1+\frac{r}{r_o}\right)^{-\gamma_o}; & \mathrm{(LT1)} \\
n_b \left ( \frac{r}{r_b} \right )^{-\gamma}\left(1+\frac{r}{r_o}\right)^{-\gamma_o}. & \mathrm{(LT2)} \\
\end{array} \right .
\label{eq:LT_models}
\end{align}
Here, the fixed parameters are $\gamma_i=0.7$, $\gamma=1.8$, $\gamma_o=1.2$, $\alpha=4$, $r_b=0.3\,\mathrm{pc}$ and $r_o = 5 \, \mathrm{pc}$. The values of the parameters $\gamma_i$, $\gamma$, $\alpha$ are adopted from \citet{merritt10}, who fitted the surface density as a function of the projected radius to the number counts of late-type stars in the sample of \citet{bse09} (we adopt $\gamma_i=0.7$ rather than $\gamma_i=0.5$ for computational reasons). For LT1, and at small radii the number density has a slope of $-\gamma_i$, which turns over to a slope of $-\gamma$ at intermediate radii; the parameter $\alpha$ determines the smoothness of this transition. For both LT1 and LT2 and at larger radii $r\gtrsim r_o$ the slope $\approx-(\gamma+\gamma_o)=-3$, which is consistent with the observed slope at these larger radii \citep{schodel_ea14}. 

As shown by \citet{merritt10} the break radius $r_b$ may have been larger in the past. In particular, the present-day core can be explained by assuming an initial core of size $r_b \sim 1 \, \mathrm{pc}$. We will show in Section~\ref{sect:dyn_ff_ast} that our results of the disruption rates at 10 Gyr vary little between models LT1 (with a core) and LT2 (without a core), indicating that the details of the core are not important for the purposes of this work.

We model the late-type stars as a single-mass stellar population with $m_\mathrm{LT} = 1 \, \mathrm{M_\odot}$. The normalization $n_b$ in equation (\ref{eq:LT_models}) is determined by equating the enclosed late-type stellar mass implied by this equation at a reference radius of $r_0 \equiv 1 \, \mathrm{pc}$, to the inferred distributed mass within $r_0$ of $M_0 \approx 1.5 \times 10^6 \, \mathrm{M_\odot}$ as determined by \citet{schodel_ea09}. Here, we neglect the contribution of early-type stars and stellar black holes (discussed below). The latter populations have a total mass of $\sim 10^4 \, \mathrm{M_\odot}$, which can safely be neglected compared to the total mass of the $\sim 10^6$ late-type stars within $r_0$. We subsequently find $n_b \approx 1.7 \times 10^6 \, \mathrm{pc^{-3}}$ and $n_b \approx 1.4 \times 10^6 \, \mathrm{pc^{-3}}$ for LT1 and LT2, respectively. 

The (negative of the) gravitational potential $\psi(r)$ is required for the calculations presented below. It is computed from equation~(\ref{eq:LT_models}) using the inverted Poisson equation (e.g. \citealt{cohn79}, \citealt[][3.2]{bookmerritt13})
\begin{align}
\nonumber \psi(r) &= \frac{GM_\bullet}{r} + \frac{4\pi G m_\mathrm{LT}}{r} \int_0^r \mathrm{d} r' \, r'^2 n_\mathrm{LT}(r') \\
&\quad - 4 \pi G m_\mathrm{LT} \int_0^r \mathrm{d} r' \, r' n_\mathrm{LT}(r').
\label{eq:psi_r}
\end{align} 
Here, the contributions of early-type stars, stellar black holes and massive perturbers (see below) to the potential are neglected. In addition we define the arbitrary constant in the potential such that $\psi(r)\rightarrow GM_\bullet/r$ as $r\rightarrow0$ assuming a power-law dependence of $n_\mathrm{LT}(r)$ with $\gamma<2$ for small $r$, which is consistent with equation~(\ref{eq:LT_models}). 

\subsubsection{Early-type stars}
\label{sect:models:GC:ET}
In addition to the late-type stars, $\sim 10^2$ young massive stars are observed within the central parsec of the GC, of which a fraction of $\sim 0.2$ is distributed within at least one disc structure with an inner edge at projected radius $\approx 0.8''$ or $\approx 32 \, \mathrm{mpc}$ \citep{paumard_ea06,bartko_ea09,lu_ea09}. The age of these stars is $\sim 2-6 \, \mathrm{Myr}$, and evidence exists for multiple formation events in the past, with an interval of $\sim 100 \, \mathrm{Myr}$ \citep{blum_ea03}. Within the central arcsecond ($\lesssim 0.04 \, \mathrm{pc}$), $\sim 20$ less massive ($3\lesssim m/\mathrm{M_\odot} \lesssim 15$) stars have been observed \citep{genzel_ea03b,eisenhauer_ea05,ghez_ea08,gillessen_ea09}. These stars, the S-stars, are likely older than $2-6 \, \mathrm{Myr}$, and their formation process so close to the SBH is unclear, although disruption of a stellar binary through the Hills mechanism \citep{hills88} is currently favoured (see \citealt{alexander05} and \citealt{geg10} for reviews).

Their short life time, relatively low total mass ($\sim 10^4 \, \mathrm{M}_\odot$) and limited radial extent make it unlikely that early-type stars are dynamically important for most planetesimals in the GC. Nevertheless, debris discs around early-type stars could be an important source of planetesimals on much longer time-scales ($>\mathrm{Gyr}$) considering that planetesimals, once stripped by the tidal force of the SBH (cf. Section~\ref{sect:stripping:SBH}), would accumulate over time while early-type stars are formed episodically. 

We estimate the latter contribution by the following source term (cf. equation~\ref{eq:FP_main})
\begin{align}
F_\mathrm{strip,ET}(\Ec) = N_\mathrm{a/\star} N_\mathrm{ET}(\Ec) / \tau_\mathrm{SF}.
\label{eq:F_strip_ET}
\end{align}
Here, $N_\mathrm{ET}(\Ec)$ is the number of early-type stars with orbital energies between $\Ec$ and $\Ec + \mathrm{d}\Ec$, where the orbital energy $\Ec = -v^2 + \psi(r)$ is defined with respect to the SBH (note that the early-type stellar mass is not included in $\psi(r)$). The quantity $\tau_\mathrm{SF}=100\,\mathrm{Myr}$ is an estimate of the time-scale at which the early-type stars are formed. In equation~(\ref{eq:F_strip_ET}) it is assumed that all planetesimals are stripped within the life time of the early-type stars, which is justified considering the narrow radial extent of the early-type stars. Because likely not all planetesimals can be stripped within the lifetime of the early-type stars, equation~(\ref{eq:F_strip_ET}) provides an upper limit for the stripping flux. Furthermore it is assumed in equation~(\ref{eq:F_strip_ET}) that over longer time-scales $\gg \tau_\mathrm{SF}$, the source term can be considered constant.

To compute $N_\mathrm{ET}(\Ec)$ we assume that the surface density of the early-type stars is $\Sigma_\mathrm{ET}(r) \propto r^{-n}$ with $n=2$ (e.g. \citealt{bartko_ea09}), which implies $N_\mathrm{ET} \propto \Ec^{n-3}$. Subsequently, $N_\mathrm{ET}(\Ec)$ is normalized by setting the total number of early-type stars formed in each formation event equal to $100$ (roughly the current observed number) and assuming a distribution in energy bounded by $\Ec_1<\Ec<\Ec_2$, where $\Ec_i=GM_\bullet/(2a_i)$, and $a_1=0.4\,\mathrm{pc}$ and $a_2=0.04\,\mathrm{pc}$.

\subsubsection{Stellar black holes}
\label{sect:models:GC:stellar_black_holes}
A population of stellar black holes distributed in a cusp close to the SBH could be dynamically important for planetesimals in the GC. Although so far not directly supported by observations, there are various theoretical motivations for the existence of such a population:

(1) Theoretical models of the GC based on the Fokker-Planck equation predict a dense population of stellar black holes close to the SBH arising from mass seggregation \citep{ha06}. 

(2) In the case of formation of the S-stars through the Hills mechanism the predicted eccentricities are on average higher than the observed eccentrities of the S-stars. Dynamical evolution by field stars through resonant relaxation (RR) can modify the high eccentricities to a distribution that is consistent with observations. When assuming a core of late-type stars the time-scale of this relaxation process is much longer than the age of the S-stars, however. This problem can be remedied when assuming a cusp of stellar black holes; in this case the relaxation time-scale is consistent with the age of the S-stars \citep{perets_ea_09}, in particular if relativistic effects are taken into account \citep{am13,antonini14,hpm14}. 

(3) If $\sim 10-100$ early-type stars in the GC are produced every $\sim 100 \, \mathrm{Myr}$ and each produce a stellar black hole $\sim 10 \, \mathrm{M_\odot}$, then a black hole population with a total mass of $\sim10^4-10^5 \, \mathrm{M_\odot}$ could be produced over a time span of 10 Gyr (e.g. \citealt{merritt10}).

We model a cusp of stellar black holes by assuming a total of 4800 stellar black holes, each $10 \, \mathrm{M}_\odot$, distributed with an approximate number density $n(r)\propto r^{-7/4}$ within $r_d\equiv 0.2 \, \mathrm{pc}$ of the SBH, similar to what was assumed in \citet{am13}. We assume that the stellar black hole distribution function vanishes for $\Ec<\Ec_d$, where $\Ec_d\equiv GM_\bullet/(2r_d)$, and neglect the effect of the black holes on the potential $\psi(r)$. We also include the possibility of a time dependence of the black hole density as implied by argument (3) above by assuming, for simplicity, that the black hole density increases linearly with time from 0 at $t=0$ to the maximum density corresponding to $N_\mathrm{BH} = 4800$ at $t=10\,\mathrm{Gyr}$. 

\subsubsection{Massive perturbers}
\label{sect:models:GC:MP}
Massive perturbers such as giant molecular clouds can strongly enhance the rate of relaxation in galactic nuclei \citep{pha07} (hereafter PHA07). They may therefore be relevant for the supply of planetesimals to the loss cone of the SBH. To evaluate the potential effect on the relaxation rate we adopt the model GMC1 from PHA07 in which the effect on the relaxation rate is the largest, therefore giving an upper limit for this effect. We take into account the effect of the massive perturbers by multiplying the inverse relaxation time-scale by $\mu_2$, the ratio of the second moments of the mass distributions of massive perturbers and stars, at the two radial bins included in table 2 of PHA07. Here, we neglect the smaller Coulomb factor in the case of massive perturbers.

\subsection{Disc models}
\label{sect:models:disc_models}
In our second formation scenario for planetesimals in galactic nuclei it is assumed that they are born in debris discs around stars (cf. Section~\ref{sect:introduction}). We assume a thin disc with surface density $\Sigma(d) \propto d^{-\beta}$, where $d$ is the distance from the star, and adopt $\beta = 1.5$ \citep{hayashi81}. The initial radial extent of the disc is $d_1 < d < d_2$, where two radial distributions are adopted for the disc as in \citet{lmlp11}: (1) $d_1 = 40 \, \mathrm{AU}$ and $d_2 = 100 \, \mathrm{AU}$ (disc model 1; `DM1') and (2) the more tightly bound distribution $d_1 = 10 \, \mathrm{AU}$ and $d_2 = 40 \, \mathrm{AU}$ (disc model 2; `DM2'). These two choices reflect uncertainties in the dependence of the radial extent of the disc on the mass of the parent star. 

Based on the bulk energy and assuming a spherical shape, ZNM12 estimated that a planetesimal of radius $\gtrsim R_\mathrm{a} \approx 10 \, \mathrm{km}$ could produce an observable flare ($L_\mathrm{X} \sim 10^{34} \, \mathrm{erg/s}$) when tidally disrupted by the SBH. Assuming a distribution of radii $\mathrm{d}N_\mathrm{a}/\mathrm{d}R_\mathrm{a} \propto R_\mathrm{a}^q$ with $q\approx -3.5$ \citep{wyatt08} and density $\rho_\mathrm{a}=1 \,\mathrm{g\,cm^{-3}}$, the number of planetesimals per star with radius $\geq R_\mathrm{a}$ is $N_\mathrm{a/\star} \sim 2 \times 10^7 [m_\mathrm{a,tot}/(10^{-5}\,\mathrm{M}_\odot)]$, where $m_\mathrm{a,tot}$ is the total mass in planetesimals per star (ZNM12). Note that $m_\mathrm{a,tot}$ is the most uncertain parameter in $N_\mathrm{a/\star}$. Below we use the term `planetesimals' to refer to planetesimals with radius $\geq R_\mathrm{a}$. 

We define the stripping radius $d_\mathrm{strip}$ as the maximum distance from a star for which a planetesimal, treated as a massless particle, is still bound to this star. The fraction $f_\mathrm{strip}$ of planetesimals that is stripped from a single star for a given $d_\mathrm{strip}$ and disc model (i.e. $d_1$, $d_2$ and $\beta$) is given by
\begin{align}
f_\mathrm{strip}(d_\mathrm{strip}) = \left \{
\begin{array}{ll}
\displaystyle 1, &d_\mathrm{strip} < d_1; \\
\displaystyle \frac{ d_2^{2-\beta} - d_\mathrm{strip}^{2-\beta} }{ d_2^{2-\beta} - d_1^{2-\beta} }, & d_1 \leq d_\mathrm{strip} \leq d_2; \\
\displaystyle 0, & d_\mathrm{strip} > d_2, \\ 
\end{array}
\right.
\label{eq:fstrip}
\end{align}
and the number of stripped planetesimals per star is $f_\mathrm{strip} N_\mathrm{a/\star}$. In Section~\ref{sect:stripping} we consider two causes for stripping: the tidal force of the SBH (Section~\ref{sect:stripping:SBH}) and gravitational encounters with other stars (Section~\ref{sect:stripping:enc}).

\section{Stripping planetesimals from stars in the GC}
\label{sect:stripping}

\begin{figure}
\center
\includegraphics[scale = 0.435, trim = 0mm 0mm 0mm 0mm]{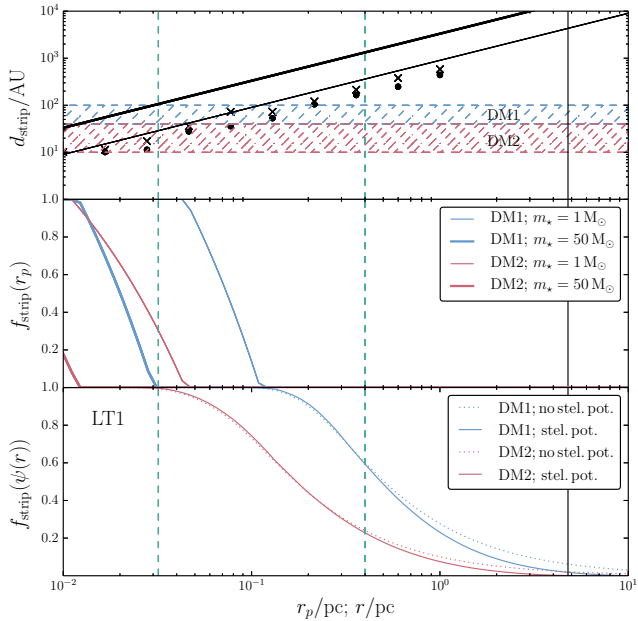}
\caption{\small {\bf Top panel}: the distance $d_\mathrm{strip}$ for which planetesimals are stripped from their parent star by the tidal force of the SBH as a function of $r_p$, the pericentre distance of the orbit of the star around the SBH, according to equation (\ref{eq:rstripSBH}). Thin black line: assuming $m_\star = 1 \, \mathrm{M_\odot}$ (i.e. late-type stars); thick black line: assuming $m_\star = 50 \, \mathrm{M_\odot}$ (i.e. early-type stars). The horizontal regions indicate the two radial extents of planetesimals around stars considered in this paper (cf. Section~\ref{sect:models:disc_models}): DM1 (blue) and DM2 (red). The green dashed vertical lines  indicate the approximate region of the young massive stars in the GC. Black bullets (crosses): stripping radii determined from $N$-body simulations assuming $e=0.01$ ($e=0.9$), see the text. {\bf Middle panel}: the stripping fraction as a function of $r_p$, computed from $d_\mathrm{strip}$ using equation~(\ref{eq:fstrip}). Blue (red) lines apply to DM1 (DM2). Thin lines: assuming $m_\star = 1 \, \mathrm{M_\odot}$; thick lines: assuming $m_\star = 50 \, \mathrm{M_\odot}$. {\bf Bottom panel}: the angular-momentum-averaged stripping fraction (assuming an isotropic velocity distribution) as a function of $r=\psi^{-1}(\Ec)$, assuming LT1. Solid lines: including the stellar potential; dashed lines: excluding the stellar potential (cf. Appendix \ref{app:fstrip:SBH}). In all panels the black solid vertical line indicates the radius of influence $r_h$ assuming LT1, defined as $M_\mathrm{LT}(r_h)=2M_\bullet$ (cf. equation~\ref{eq:M_LT}).}
\label{fig:stripping_fraction_SBH}
\end{figure}

\subsection{Stripping by the SBH}
\label{sect:stripping:SBH}
We determine $d_\mathrm{strip}$ (cf. Section~\ref{sect:models:disc_models}) for planetesimals that are initially bound to a star with mass $m_\star$, that in turn is bound to the SBH in an orbit with pericentre distance $r_p$. A simple estimate of $d_\mathrm{strip}$ by the tidal force of the SBH is given by the radius of the Hill sphere (e.g. \citealt{hb92})
\begin{align}
d_\mathrm{strip} \approx r_p \left ( \frac{m_\star}{3M_\bullet} \right )^{1/3}.
\label{eq:rstripSBH}
\end{align}

In the top panel of Fig. \ref{fig:stripping_fraction_SBH} the stripping radii according to equation~(\ref{eq:rstripSBH}) are plotted as a function of $r_p$. Two masses are adopted for the parent star: $1\,\mathrm{M_\odot}$ (thin line) and $50\,\mathrm{M_\odot}$ (thick line). The radial extents of the disc models DM1 and DM2 are indicated with the blue and red regions, respectively; these extents are used to compute the stripping fractions $f_\mathrm{strip}(r_p)$ (cf. equation~\ref{eq:fstrip}), and are shown as a function of $r_p$ in the middle panel of Fig. \ref{fig:stripping_fraction_SBH}. For both stellar masses and disc models the SBH is ineffective at stripping planetesimals from stars with pericentre distances $\gtrsim 0.1 \, \mathrm{pc}$. There is a strong dependence of $f_\mathrm{strip}(r_p)$ on both the disc models and the mass of the parent star. For instance, the pericentre distance for which $f_\mathrm{strip}=0$ is $\approx 2.5$ times larger for DM1 compared to DM2 (assuming $m_\star=1\,\mathrm{M}_\odot$). The qualitative behaviour is intuitively easy to understand: in DM2 the planetesimals are more tightly bound to the parent star, therefore the parent star needs to be closer to the SBH for the planetesimals to be stripped. Similarly, more massive parent stars also need to be closer to the SBH for effective stripping of planetesimals by the tidal force of the SBH. 

As a verification of equation~(\ref{eq:rstripSBH}) we carried out a series of $N$-body simulations of a star with mass $1\,\mathrm{M}_\odot$ orbiting the SBH. The star is orbited by a debris disc consisting of 100 particles of mass $2.0\times10^{-15}\,\mathrm{M}_\odot$ in circular, coplanar orbits and with semimajor axes ranging between 10 and 1000 AU. The pericentre distance of the stellar orbit $r_p$ is varied between 0.01 and 1 pc. The eccentricity $e$ of the latter orbit is assumed to be either 0.01 or 0.9 and the semimajor axis is computed from $a=r_p/(1-e)$. The system is integrated for the duration of an orbit of the star around the SBH with the \textsc{hermite0} code \citep{hmm95} in the \textsc{AMUSE} framework \citep{pelupessy_ea13,pmepv13}. 

From each $N$-body simulation we determined the orbital elements of the debris disc particles with respect to the star, and recorded which particles become unbound from the star at some point in the integration. For the latter particles, the minimum of the initial pericentre distances with respect to the star was adopted as the stripping radius $d_\mathrm{strip}$.

In the top panel of Fig. \ref{fig:stripping_fraction_SBH} the stripping radii inferred from the $N$-body simulations are shown with the black bullets (crosses) assuming the stellar orbit has $e=0.01$ ($e=0.9$). The stripping radii determined from the $N$-body simulations are slightly lower compared to those implied by equation~(\ref{eq:rstripSBH}); the discrepancy is largest for the nearly circular orbit (black bullets). Nevertheless, the discrepancy, averaging the stripping radii from the $N$-body simulations for the two cases $e=0.01$ and $e=0.9$, is no larger than a factor of $\approx 1.9$. Therefore, we believe equation~(\ref{eq:rstripSBH}) is adequate for the purposes of this paper. 

In equation~\ref{eq:rstripSBH} the stripping fraction $f_\mathrm{strip}$ is expressed in terms of the pericentre distance of the star. For the purposes of Section~\ref{sect:dyn_ff_ast} we also compute $f_\mathrm{strip}$ in terms of the orbital energy $\mathcal{E}$ assuming an isotropic velocity distribution, and taking into account the stellar potential. The details are included in Appendix \ref{app:fstrip:SBH}. 

In the bottom panel of Fig. \ref{fig:stripping_fraction_SBH} the quantity $f_\mathrm{strip}(\mathcal{E})$ is plotted as a function of radius by setting $\Ec=\psi(r)$, i.e. $r=\psi^{-1}(\Ec)$ where $\psi^{-1}(\Ec)$ is the inverse function of $\psi(r)$, and assuming late-type model LT1. Here, we include two cases: with (solid lines) and without (dotted lines) the stellar potential; in the latter case analytic expressions can be derived for $f_\mathrm{strip}(\Ec)$ (cf. Appendix \ref{app:fstrip:SBH}). As expected, at small $r$ it is a good approximation to neglect the stellar potential; at larger $r$, however, neglect of the stellar potential causes the stripping fraction to be slightly overestimated. 

Compared to the case if $f_\mathrm{strip}$ is expressed in terms of the pericentre distance $r_p$ (cf. the middle panel of Fig. \ref{fig:stripping_fraction_SBH}), if expressed in terms of orbital energy (cf. the third panel of Fig. \ref{fig:stripping_fraction_SBH}) it is non-zero for a much larger radial range. This is because for a given $r_p$ there exist many (eccentric) orbits with typical radius $r\gg r_p$.

\subsection{Stripping by gravitational encounters with other stars}
\label{sect:stripping:enc}
A planetesimal bound to a star orbiting the SBH can be treated as a binary system orbiting the SBH. Because of gravitational encounters with other stars the intrinsic binary parameters, in particular the semimajor axis $a_\mathrm{bin}$, change over time. A quantity that describes whether $a_\mathrm{bin}$ on average increases or decreases is the hardness parameter $h$, the ratio of the (negative) specific binding energy of the binary, $\Ec_\mathrm{bin}=Gm_\mathrm{bin}/(2a_\mathrm{bin})$, to the squared stellar velocity dispersion, $\sigma^2(r)$ \citep{heggie75}. 
Here, $m_\mathrm{bin} \equiv m_\star + m_\mathrm{a} \approx m_\star$. If $h\ll1$ the binary is soft and gravitational encounters on average cause such binaries to become softer (i.e. to decrease $h$), until they dissolve as $h\rightarrow0$ (e.g. \citealt{heggie75,hut83,hutbahcall83}). On the other hand, if $h\gg1$ the binary is hard and encounters tend to make it even harder (e.g. \citealt{heggie75,hut93,heggiehut96}). 

\begin{figure}
\center
\includegraphics[scale = 0.435, trim = 0mm 0mm 0mm 0mm]{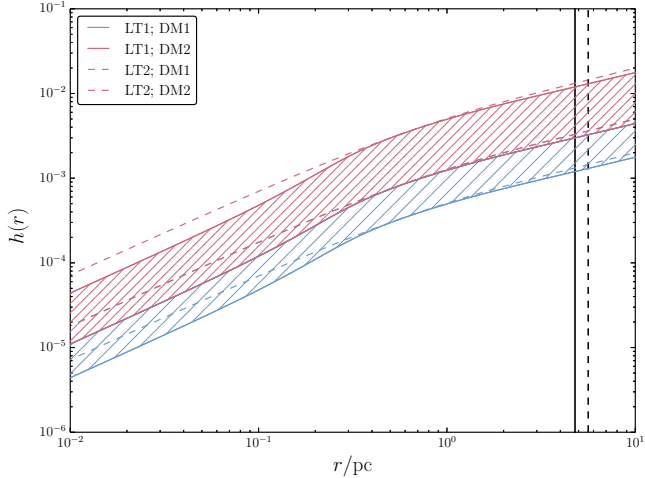}
\caption{\small The hardness parameter $h$ as a function of the distance $r$ to the SBH for planetesimals distributed around the star according to DM1 (blue regions) and DM2 (red regions). The velocity dispersion is given by equation~(\ref{eq:jeans}) and $m_\mathrm{bin}=1\,\mathrm{M}_\odot$ is assumed. Solid (dashed) lines apply to LT1 (LT2). Vertical lines: radius of influence in the models LT1 (solid) and LT2 (dashed).}
\label{fig:hardness_parameter}
\end{figure}

To investigate which of these regimes applies to planetesimals in the GC we compute $\sigma(r)$ from the isotropic Jeans equation, and assume that the number density and mass are dominated by late-type stars,
\begin{align}
n_\mathrm{LT}(r) \sigma^2(r) = \int_r^\infty \mathrm{d} r' \, \frac{G M(r') n_\mathrm{LT}(r')}{{r'}^2},
\label{eq:jeans}
\end{align}
where $M(r) = M_\bullet + M_\mathrm{LT}(r)$ and
\begin{align}
M_\mathrm{LT}(r) = 4 \pi m_\mathrm{LT} \int_0^r n_\mathrm{LT}(r') r'^2 \, \mathrm{d} r'.
\label{eq:M_LT}
\end{align}
Adopting the semimajor axes from the disc models DM1 and DM2 and assuming $m_\mathrm{bin} = m_\mathrm{LT} = 1 \, \mathrm{M_\odot}$, the hardness parameters that follow from equation~(\ref{eq:jeans}) are plotted in Fig. \ref{fig:hardness_parameter} for LT1 (solid lines) and LT2 (dashed lines). For all combinations of disc models and distributions of late-type stars, $h<10^{-2}$ for $r<r_h$, which shows that in our models the planetesimal+star binary is initially always soft. Therefore, gravitational encounters on average soften these binaries even further, to the point that eventually the planetesimals are no longer bound to their parent star. 

For soft binaries the stripping time-scale, i.e. the time-scale for the orbital energy of the binary to change by order itself because of gravitational encounters, can be estimated by \citep{bookbt08,perets_09,ap13}
\begin{align}
t_\mathrm{strip} \equiv \frac{|\Ec_\mathrm{bin}|}{\langle \mathcal{D}(\Delta \Ec_\mathrm{bin}) \rangle} = \frac{1}{8} \sqrt{\frac{1+q_\sigma}{2\pi q_\sigma}} \frac{m_\mathrm{bin} \sigma(r)}{G n(r) \langle m_\star^2 \rangle a_\mathrm{bin} \log(\Lambda_\mathrm{bin})}.
\label{eq:t_strip}
\end{align}
Here, $\langle \mathcal{D}(\Delta \Ec_\mathrm{bin}) \rangle$ is a diffusion coefficient for the binary binding energy, $q_\sigma \equiv m_\mathrm{bin}/m_\star \approx 1$, $\langle m_\star^2 \rangle$ is the second moment of the stellar mass function, and $\Lambda_\mathrm{bin}$ is the Coulomb factor for evaporation in the soft limit, which can be estimated by $\Lambda_\mathrm{bin} \sim 3 (1+1/q_\sigma)/(1+2/q_\sigma) [\sigma^2(r)/v_\mathrm{bin}^2]$, where $v_\mathrm{bin}^2 \equiv Gm_\mathrm{bin}/a_\mathrm{bin}$ \citep{ap13}.

We assume that the parent star has mass $m_\mathrm{bin} \approx m_\mathrm{LT} = 1 \, \mathrm{M_\odot}$ and we consider encounters with late-type stars only, such that $\langle m_\star^2 \rangle = m_\mathrm{LT}^2$, $n(r) = n_\mathrm{LT}(r)$ and $\sigma$ is given by equation~(\ref{eq:jeans}). Evidently, encounters with early-type stars also lead to evaporation. However, at the radial positions of the majority of the early-type stars in the GC, the SBH is already effective at stripping most planetesimals (cf. Fig. \ref{fig:stripping_fraction_SBH}). Furthermore, because massive perturbers are extended objects they do not contribute to stripping (PHA07). Therefore, we expect, in the case of the GC, that stripping by encounters at larger radii is dominated by encounters with late-type stars.

\begin{figure}
\center
\includegraphics[scale = 0.435, trim = 0mm 0mm 0mm 0mm]{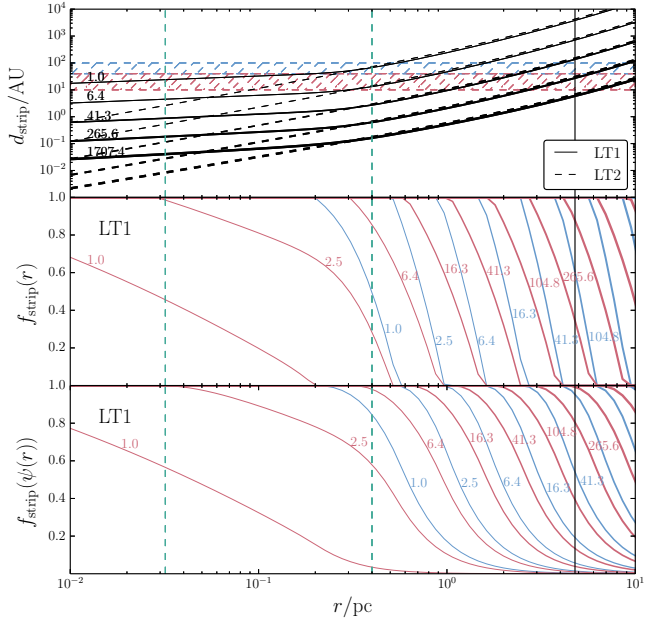}
\caption{\small {\bf Top panel}: the distance $d_\mathrm{strip}$ for which planetesimals are stripped from their parent star by gravitational encounters with other stars as a function of the distance $r$ to the SBH, computed from equation (\ref{eq:t_strip}) for different ages. Age increases with line thickness and is indicated (in Myr) with numbers. Black solid line: assuming LT1; black dashed line: assuming LT2; in both cases $m_\mathrm{bin}=1\,\mathrm{M}_\odot$. {\bf Middle panel}: the stripping fraction as a function of $r$ at various ages indicated in Myr with numbers, computed from $d_\mathrm{strip}$ using equation~(\ref{eq:fstrip}) and assuming LT1. Blue (red) lines apply to DM1 (DM2). {\bf Bottom panel}: the angular-momentum- and orbit-averaged stripping fraction as a function of $r=\psi^{-1}(\Ec)$ at various ages, assuming LT1. In all panels the black solid vertical line indicates the radius of influence $r_h$ for LT1. }
\label{fig:stripping_fraction_enc}
\end{figure}

Using equation~(\ref{eq:t_strip}) we estimate the stripping semimajor axis $d_\mathrm{strip}$ because of gravitational encounters with late-type stars as follows (here, we assume that the orbits of the planetesimals around the star are circular). For a given distance $r$ to the SBH and age $t$ of the star+planetesimal binary system the stripping time-scale $t_\mathrm{strip}$ is equated to $t$. The resulting equation is solved numerically for $a_\mathrm{bin}=d_\mathrm{strip}$ (note that the logarithmic term $\log(\Lambda_\mathrm{bin})$ depends on $a_\mathrm{bin}$). 

In the top panel of Fig. \ref{fig:stripping_fraction_enc} we show the resulting $d_\mathrm{strip}$ as a function of the distance $r$ of the centre of mass of the binary system, for the models LT1 (black solid lines) and LT2 (black dashed lines) and assuming different ages $t$. At small radii the stripping radius is larger for LT1 than for LT2. This can be attributed to the lower number density at small radii in the case of LT1, implying less efficient stripping.

We show in the middle panel of Fig. \ref{fig:stripping_fraction_enc} the stripping fraction as a function of $r$ at various ages, indicated with numbers, for LT1 and assuming DM1 (blue lines) and DM2 (red lines). There is a strong dependence of $f_\mathrm{strip}(r)$ on time. For instance, after $1\,\mathrm{Myr}$ and for DM1 the fraction at $0.4\,\mathrm{pc}$, $\approx 0.5$, is comparable to the fraction in the case of stripping by the tidal force of the SBH alone (cf. the bottom panel of Fig. \ref{fig:stripping_fraction_SBH}), whereas after $\sim 100 \, \mathrm{Myr}$ stripping has crept up all the way to the radius of influence. 

In the above the stripping fraction $f_\mathrm{strip}$ was computed as a function of $r$. For a typical orbit around the SBH $r$ is not constant, however: the star will pass through environments with different densities and velocity dispersions. To take this into account we performed an orbital average of $f_\mathrm{strip}(r)$ weighted according to the time spent at each point in the orbit, and assuming an isotropic velocity distribution. See Appendix \ref{app:fstrip:enc} for details. 

In the bottom panel of Fig. \ref{fig:stripping_fraction_enc} the orbit-averaged stripping fraction, $f_\mathrm{strip}(\Ec)$, is plotted as a function of $r=\psi^{-1}(\Ec)$ for various ages and assuming LT1. Compared to the non-averaged $f_\mathrm{strip}(r)$ (cf. the second panel of Fig. \ref{fig:stripping_fraction_enc}), the orbit-averaged stripping fractions are somewhat `smeared out' to larger radii. In other words, allowing the orbital radius to vary during the orbit results in stripping by encounters to larger distances from the SBH.

\section{Dynamics of planetesimals orbiting the SBH}
\label{sect:dyn_ff_ast}
\subsection{Fokker-Planck equation}
\label{sect:dyn_ff_ast:eq}
Planetesimals orbiting the SBH are susceptible to gravitational encounters with stars and other massive objects. The end result of this process is a steady-state in orbital angular momentum and energy. Previous studies (e.g. \citealt{bw77,merritt04,mhb07,merritt10}; \citealt[][7.1.2.2]{bookmerritt13}) have shown that in the case of scattering of a light population (i.e. planetesimals) by a massive population, the steady-state corresponds to a number density $n(r) \propto r^{-3/2}$. This is achieved on a time-scale $\sim t_\Ec$, the time-scale for non-resonant relaxation (NRR) to change the orbital energy by order itself. 

Here, we explore specifically the case of scattering of planetesimals by various perturbers in the GC. In addition we take into account other potentially important effects such as the time- and energy-dependent influx of planetesimals from stars in our second scenario of planetesimal formation (cf. Section~\ref{sect:introduction}), as well as collisions and resonant relaxation (RR). Our main focus is the disruption rate of planetesimals by the SBH as a function of time, and in particular at $t=10\,\mathrm{Gyr}$, which is approximately the age of the majority of the late-type stars in the GC (cf. Section~\ref{sect:models:GC}). 

To model the orbital evolution of planetesimals bound to the SBH we solve the time-dependent Fokker-Planck equation in energy space for gravitational scattering of (nearly) massless particles (i.e. planetesimals) by massive particles with a central SBH (e.g. \citealt[][7.1.2.2]{bookmerritt13}), including several additional terms,
\begin{align}
4 \pi^2 p(\Ec) \nonumber \frac{\partial f_\mathrm{a}(\Ec,t)}{\partial t} &= \frac{\partial \mathcal{F}_\Ec(\Ec,t)}{\partial \Ec} + F_\mathrm{strip}(\Ec,t) - F_\mathrm{lc}(\Ec,t) \\
&\quad - F_\mathrm{col}(\Ec,t) - F_\mathrm{RR}(\Ec,t).
\label{eq:FP_main}
\end{align}
Here, $\Ec=-\frac{1}{2}v^2 + \psi(r)$ is the binding energy per unit mass of the planetesimals with respect to the SBH, $f_\mathrm{a}(\Ec,t)$ is the planetesimal distribution function and $p(\Ec)$ is the phase space volume per unit energy (cf. equation~\ref{eq:p_e}). The number of planetesimals with energies between $\Ec$ and $\Ec + \mathrm{d}\Ec$ at time $t$ is $N_\mathrm{a} (\Ec,t) \, \mathrm{d}\Ec$, where $N_\mathrm{a}$ is related to the distribution function via $N_\mathrm{a}(\Ec,t) = 4\pi^2 p(\Ec) f_\mathrm{a}(\Ec,t)$. 

Equation~(\ref{eq:FP_main}) is derived under the assumption that the distribution function is (nearly) isotropic in velocity space and therefore independent of angular momentum (near the loss cone a logarithmic dependence on angular momentum is implicitly assumed, cf. Appendix \ref{app:FP_main:lc}). On time-scales $\ll t_\Ec$ the evolution is expected to be dominated by changes in angular momentum, resulting in a steady-state with respect to that quantity \citep{fr76}. The latter steady-state is subsequently expected to be maintained while the distribution in energies changes, because relaxation in energy occurs on the longer time-scale of $\sim t_\Ec$. In particular, for the GC it was shown by \citet{merritt10} that an initially anisotropic distribution of late-type stars with a core in the GC evolves to near isotropy on a time-scale of $1\,\mathrm{Gyr}$, which is much shorter than the relaxation time-scale at the radius of influence derived by \citet{merritt10}, $t_\Ec \sim 20 -30 \, \mathrm{Gyr}$. Here, we are interested in the long-term ($> \, \mathrm{Gyr}$) evolution of the planetesimals. For this purpose it is therefore justified to assume isotropy in equation~(\ref{eq:FP_main}).

The term $\partial \mathcal{F}_\Ec / \partial \Ec$ in equation~(\ref{eq:FP_main}) describes the effect of gravitational scattering of planetesimals by more massive objects. In all our models we consider scattering by late-type stars (cf. \,\S\ref{sect:models:GC:LT}). In a number of models we also consider scattering by a (hypothesized) cusp of stellar black holes (cf. Section~\ref{sect:models:GC:stellar_black_holes}) and massive perturbers (cf. Section~\ref{sect:models:GC:MP}). In each case we assume that the distribution of scatterers is constant with time. 

The terms $F_i(\Ec,t)$ in equation~(\ref{eq:FP_main}), which are all positive, represent sinks (negative sign preceding $F_i$) and sources (positive sign preceding $F_i$) of planetesimals. We consider the following effects. The source term $F_\mathrm{strip}$ describes the stripping of planetesimals from their parent stars. For late-type stars it is computed from
\begin{align}
F_\mathrm{strip,LT}(\Ec,t) = N_\mathrm{a/\star} N_\mathrm{LT}(\Ec) \frac{\partial f_\mathrm{strip,LT}(\Ec,t)}{\partial t}.
\label{eq:F_strip_LT}
\end{align}
Here, $N_\mathrm{a/\star}$ is the number of planetesimals per star, $N_\mathrm{LT}(\Ec) \, \mathrm{d} \Ec$ is the number of late-type stars with energies between $\Ec$ and $\Ec+\mathrm{d}\Ec$, and $f_\mathrm{strip,LT}(\Ec,t)$ was computed in Section~\ref{sect:stripping:enc}. In the case of stripping from early-type stars equation~(\ref{eq:F_strip_ET}) is adopted. 
We note that our approach of including a source term in the  Fokker-Planck equation in energy space to model inflow of  planetesimals bound to the SBH is similar to the approach of \citet{aharon_perets_14}, who include a source term to model the formation of  stars in nuclear star clusters.
Furthermore, the sink term $F_\mathrm{lc}$ describes the effects of gravitational encounters into the loss cone of the SBH in angular-momentum space. In addition, physical collisions are described by the sink term $F_\mathrm{col}$; we consider collisions of planetesimals with late-type stars and with other planetesimals. Lastly, the sink term $F_\mathrm{RR}$ describes the effects of resonant relaxation (RR). 

More details on the terms that appear on the right-hand side of equation~(\ref{eq:FP_main}) are included in Appendix \ref{app:FP_main}. To investigate the relative importance of the effects described by these terms we solve equation~(\ref{eq:FP_main}) for each effect separately. Our models are described in Table \ref{table:dis_rates_description}. 

In each integration of equation~(\ref{eq:FP_main}) we evaluate the (energy-integrated) disruption rate of planetesimals by the SBH, $F_\mathrm{dis}(t)$, defined by
\begin{align}
F_\mathrm{dis}(t) = \int \mathrm{d} \Ec \, \left [ F_\mathrm{lc}(\Ec,t) + F_\mathrm{RR}(\Ec,t) \right ] + \mathcal{F}_\Ec(\Ec_\mathrm{lc}).
\label{eq:F_dis}
\end{align}
Here, $\Ec_\mathrm{lc} = GM_\bullet/(2r_\mathrm{lc})$ is the energy of an orbit with semimajor axis $r_\mathrm{lc}$. The term $\mathcal{F}_\Ec(\Ec_\mathrm{lc})$ 
represents the loss cone flux arising from scattering in energy space. Although included in the calculations for completeness, we find that $\mathcal{F}_\Ec(\Ec_\mathrm{lc})$ is invariably negligible compared to the flux in angular-momentum space, which is represented by the first two terms on the right-hand side of equation~(\ref{eq:F_dis}). The largest fractional contribution of the energy flux to the total rate for the rates given in Table \ref{table:dis_rates} $\approx 2 \times 10^{-5}$.

\subsection{Boundary conditions and initial conditions}
\label{sect:dyn_ff_ast:init}
Equation~(\ref{eq:FP_main}) is integrated in time using the \textsc{Python} \textsc{FiPy} library \citep{gww09}. For each model integration 200 steps are taken in which the term $F_\mathrm{strip}(\Ec,t)$, which explicitly depends on time, is updated. We have verified that increasing this number of steps does not change the results. The orbital energies are discretized on to a grid that is constructed using the relation $\Ec = \psi(r)$, where $r$ is sampled from a logarithmic grid in the range $2\,r_\mathrm{lc}\leq r \leq 100\,\mathrm{pc}$, and $r_\mathrm{lc}=1\,\mathrm{AU}$. The largest value of $\Ec$, $\Ec_\mathrm{lc} = \psi(2r_\mathrm{lc})$, corresponds to a semimajor axis of $a=r_\mathrm{lc}$ under the assumption that $\psi(r_\mathrm{lc}) = GM_\bullet/r_\mathrm{lc}$. At this energy planetesimals are considered disrupted by the SBH, and we impose the boundary condition $f_\mathrm{a}(\Ec_\mathrm{lc},t) = 0$. 

The initial distribution in the case of planetesimals initially in discs around stars, is
\begin{align}
N_\mathrm{a}(\Ec,0) = N_\mathrm{a/\star} f_\mathrm{strip;SBH}(\Ec) N_\mathrm{LT}(\Ec).
\label{eq:finit_disc}
\end{align}
Here, $f_\mathrm{strip;SBH}(\Ec)$ is the stripping fraction due to the tidal force of the SBH only (cf. Section~\ref{sect:stripping:SBH}), and $N_\mathrm{LT}(\Ec) = 4 \pi^2 p(\Ec) f_\mathrm{LT}(\Ec)$ is number of late-type stars with energies between $\Ec$ and $\Ec+\mathrm{d}\Ec$. In equation~(\ref{eq:finit_disc}) we assume that any stripping by the SBH acts instantaneously. We justify this by noting that most of the stripping by the SBH occurs at radii $\lesssim 1 \, \mathrm{pc}$ (cf. the bottom panel of Fig. \ref{fig:stripping_fraction_SBH}), for which the radial orbital period $\lesssim 7\times 10^4\,\mathrm{yr}$. We expect stripping by the SBH to occur on the latter time-scale, therefore stripping acts essentially instantaneously compared to our integration time.  

In the case of planetesimals formed in a large-scale cloud around the SBH the initial distribution is unclear. Here, we assume that the distribution is similar to the (late-type) stellar distribution, i.e. $N_\mathrm{a}(\Ec,0) \propto N_\mathrm{LT}(\Ec)$. The constant of proportionality is unconstrained, but here we assume it is given by the number of planetesimals per star, $N_\mathrm{a/\star}=2\times 10^7$, i.e. we assume that the initial distribution is
\begin{align}
N_\mathrm{a}(\Ec,0) = N_\mathrm{a/\star} N_\mathrm{LT}(\Ec).
\label{eq:finit_cloud}
\end{align}

\begin{table}
\begin{tabular}{lcccccc}
\toprule
model & \multicolumn{6}{c}{$F_\mathrm{dis}(t=10\,\mathrm{Gyr})/\mathrm{day^{-1}}$} \\
\midrule
& \multicolumn{3}{c}{LT1} & \multicolumn{3}{c}{LT2} \\
\midrule
 & cloud & \multicolumn{2}{c}{disc} & cloud & \multicolumn{2}{c}{disc} \\
& & DM1 & DM2 & & DM1 & DM2 \\
\toprule
1 & 0.55 & 0.54 & 0.53 & 0.61 & 0.59 & 0.57 \\
2 & 0.55 & 0.54 & 0.53 & 0.61 & 0.59 & 0.57 \\
3 & 0.38 & 0.37 & 0.36 & 0.51 & 0.49 & 0.47 \\
4 & 0.38 & 0.37 & 0.36 & 0.51 & 0.49 & 0.47 \\
5 & 0.55 & 0.54 & 0.53 & 0.59 & 0.57 & 0.54 \\
6 & 0.54 & 0.53 & 0.52 & 0.59 & 0.57 & 0.55 \\
7 & 0.55 & 0.54 & 0.53 & 0.61 & 0.59 & 0.56 \\
8 & 0.58 & 0.57 & 0.56 & 0.63 & 0.61 & 0.58 \\
9 & 0.55 & 0.54 & 0.53 & 0.61 & 0.59 & 0.57 \\
10 & 0.71 & 0.70 & 0.69 & 0.72 & 0.69 & 0.67 \\
11 & 0.58 & 0.57 & 0.56 & 0.63 & 0.60 & 0.58 \\
12 & 0.71 & 0.70 & 0.69 & 0.72 & 0.69 & 0.67 \\
13 & 0.58 & 0.57 & 0.56 & 0.63 & 0.60 & 0.58 \\
14 & 0.24 & 0.18 & 0.14 & 0.42 & 0.31 & 0.23 \\
\bottomrule
\end{tabular}
\caption{ Disruption rates of planetesimals with radius $\geq 10\,\mathrm{km}$ passing the SBH in the GC at distances closer than $1\,\mathrm{AU}$, possibly resulting in an observable NIR/X-ray flare (ZNM12), based on time-integrations of equation~(\ref{eq:FP_main}). The initial number of these planetesimals per star is assumed to be $N_\mathrm{a/\star}=2\times10^7$, consistent with observations of debris discs of stars in the Solar neighbourhood. Rates  at the end of our integrations, $t=10\,\mathrm{Gyr}$, are listed for various models, given in the first column, and for different combinations of the assumed distribution of late-type stars (LT1 or LT2, cf. Section~\ref{sect:models:GC:LT}). The planetesimals are assumed to be initially either distributed in a large-scale cloud around the SBH (`cloud') or in discs around stars (`disc'; the two disc models DM1 and DM2 are defined in Section~\ref{sect:models:disc_models}). Descriptions of the various models are included in Table \ref{table:dis_rates_description}. }
\label{table:dis_rates}
\end{table}

\begin{table}
\begin{tabular}{ll}
\toprule
model & description \\
\toprule
1 & LT scattering \\
2 & model 1 + source term: planetesimals from ET stars \\
3 & model 1 + BH cusp (no RR) \\
4 & model 1 + BH cusp (no RR); $t$-dependence \\
5 & model 1 + LT-planetesimal collisions ($m_\mathrm{LT,init}=1\, \mathrm{M}_\odot$) \\
6 & model 1 + LT-planetesimal collisions ($m_\mathrm{LT,init}=2\,\mathrm{M}_\odot$) \\
7 & model 1 + planetesimal-planetesimal collisions \\
8 & model 1 + RR; LT ($\chi_\mathrm{RR}=1$) \\
9 & model 1 + RR; LT ($\chi_\mathrm{RR}=0.1$) \\
10 & model 1 + RR; BH cusp ($\chi_\mathrm{RR}=1$) \\
11 & model 1 + RR; BH cusp ($\chi_\mathrm{RR}=0.1$) \\
12 & model 1 + RR; BH cusp ($\chi_\mathrm{RR}=1$; $t$-dependence) \\
13 & model 1 + RR; BH cusp ($\chi_\mathrm{RR}=0.1$; $t$-dependence) \\
14 & model 1 + massive perturbers (PHA07: GMC1) \\
\bottomrule
\end{tabular}
\caption{ Brief descriptions of the various models included in Table \ref{table:dis_rates}. Refer to the text in Section~\ref{sect:dyn_ff_ast:results} and Appendix \ref{app:FP_main} for more details. }
\label{table:dis_rates_description}
\end{table}

\begin{figure}
\center
\includegraphics[scale = 0.435, trim = 0mm 0mm 0mm 0mm]{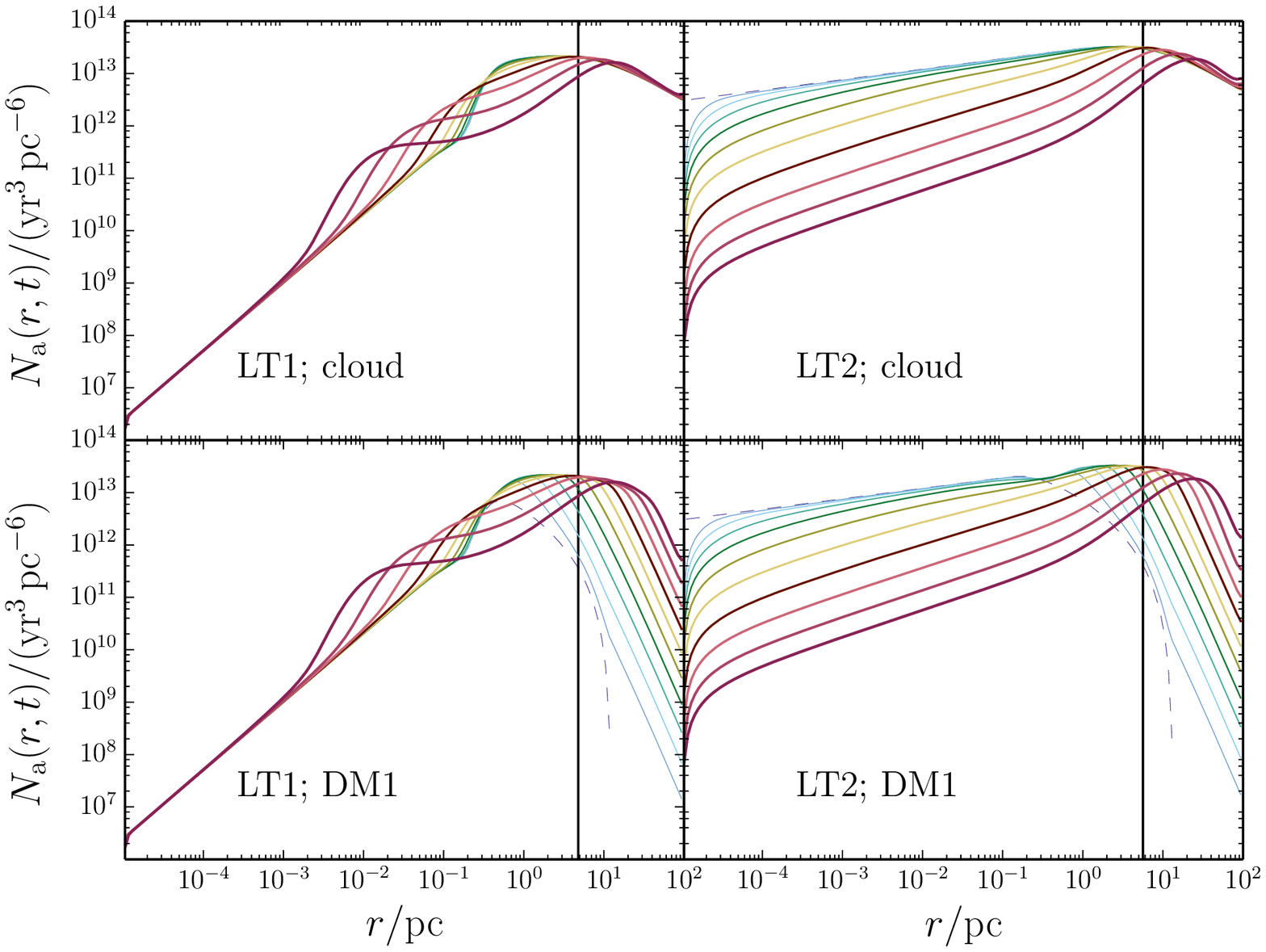}
\includegraphics[scale = 0.435, trim = 0mm 0mm 0mm 0mm]{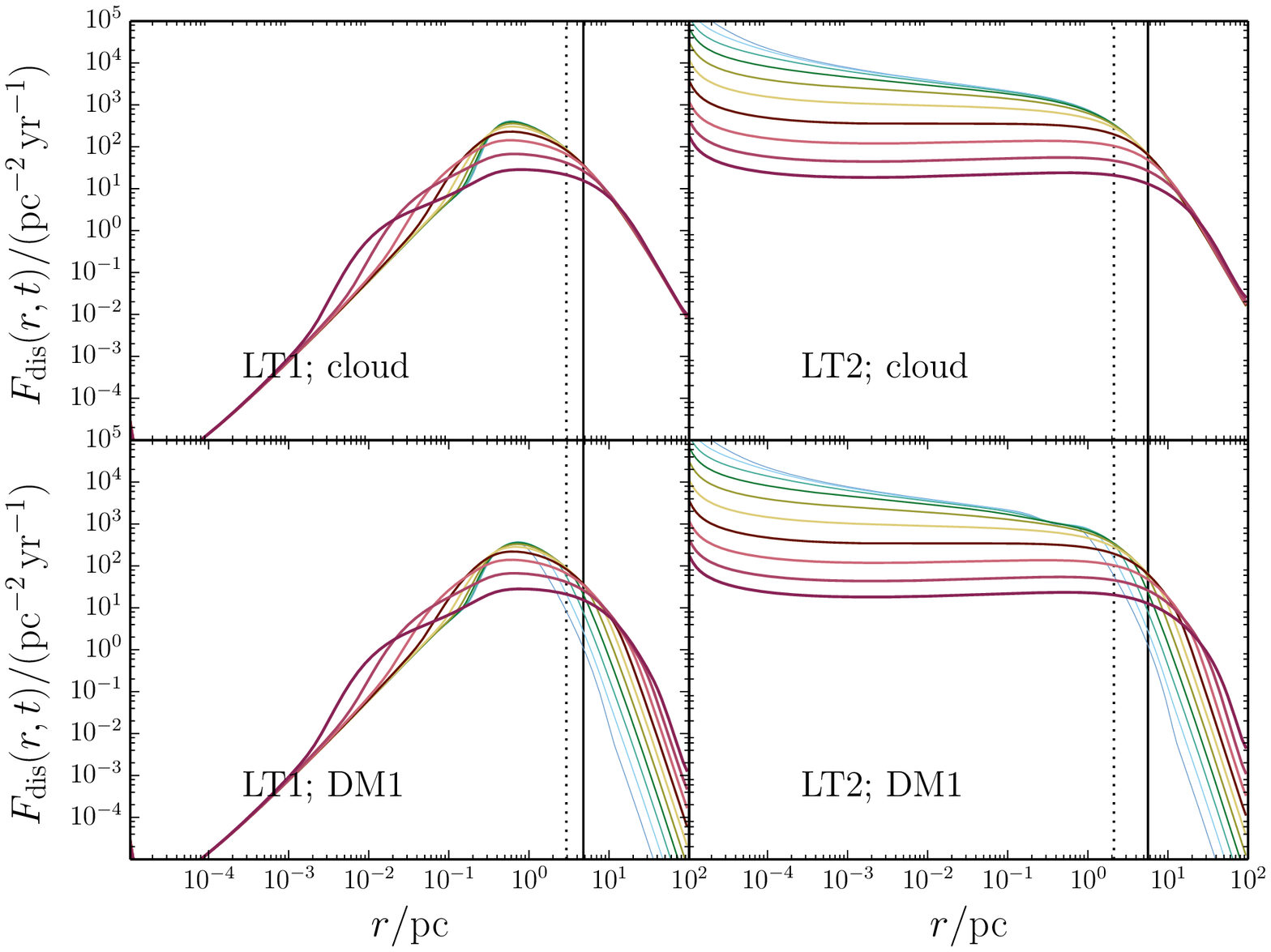}
\caption{\small {\bf Top four panels}: the quantity $N_\mathrm{a}(r,t)$, defined such that $N_\mathrm{a}(r,t)\,\mathrm{d}r$ is the number of planetesimals between radii $r$ and $r+\mathrm{d}r$, as a function of $r$ at various times, according to numerical integrations of equation~(\ref{eq:FP_main}) and assuming model 1 (cf. Table \ref{table:dis_rates_description}). The line thickness increases with time; the times shown are 0 (blue dashed line), 2.5, 6.4, 16.3, 41.3, 104.8, 265.6, 673.4, 1707.4, 4328.8 and 10000.0 Myr. The first (second) column applies to LT1 (LT2); the first (second) row applies to formation in a cloud (disc; DM1). The radii of influence $r_h$ for LT1 and LT2 are indicated with the black vertical solid lines. {\bf Bottom four panels}: the disruption flux of planetesimals by the SBH, $F_\mathrm{dis}(r,t)$ (cf. Section~\ref{sect:dyn_ff_ast:results:basic}), as a function of $r$ at various times (the meaning of the line thicknesses and colors is the same as in the top four panels). The black vertical dotted line shows the boundary between the empty and full loss cone regimes, here defined as $\tilde{q}=1$ (cf. Section~\ref{app:FP_main:lc}). }
\label{fig:N_ast_r_F_lc_r_main_models}
\end{figure}

\begin{figure}
\center
\includegraphics[scale = 0.435, trim = 0mm 0mm 0mm 0mm]{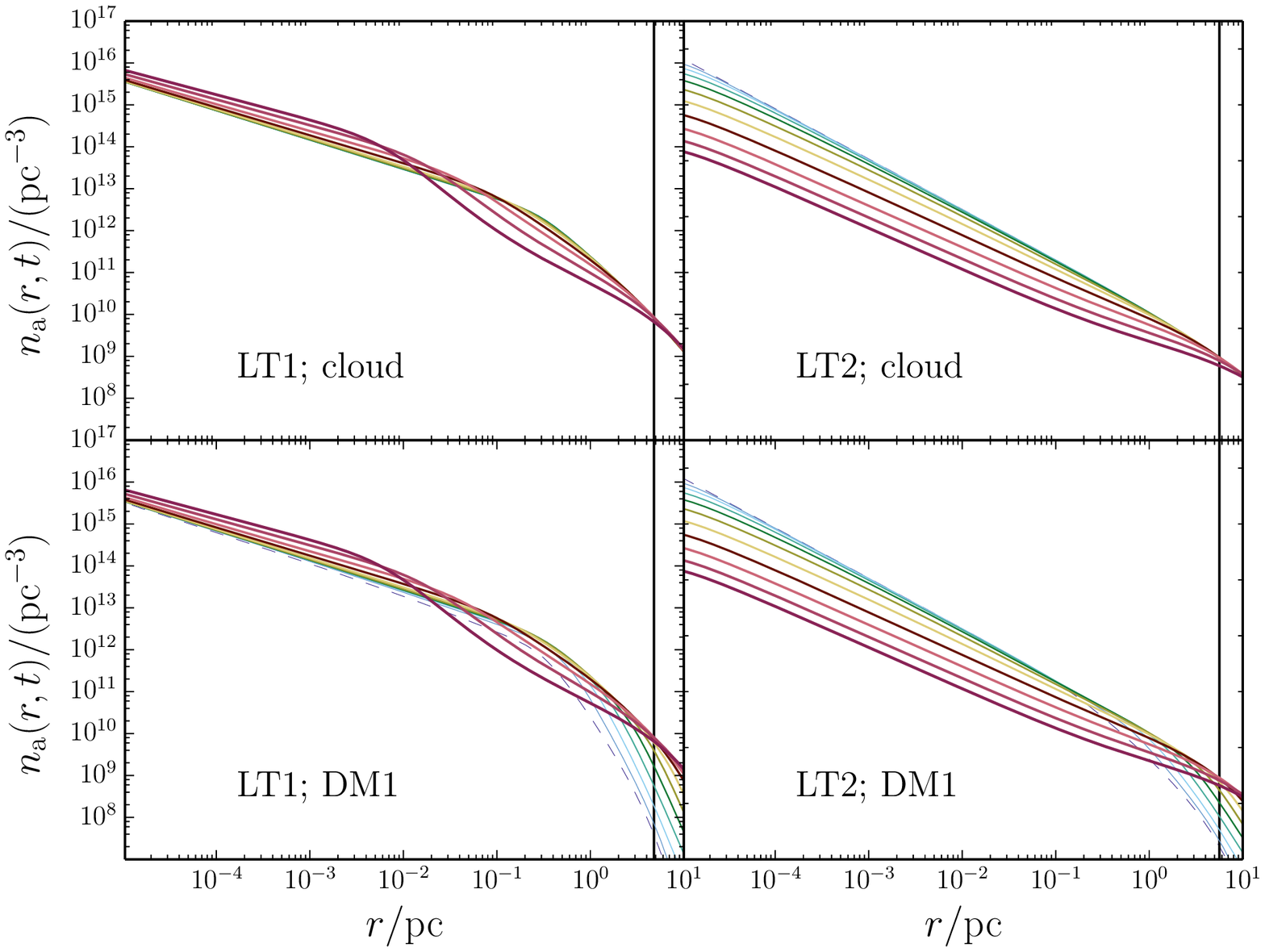}
\includegraphics[scale = 0.435, trim = 0mm 0mm 0mm 0mm]{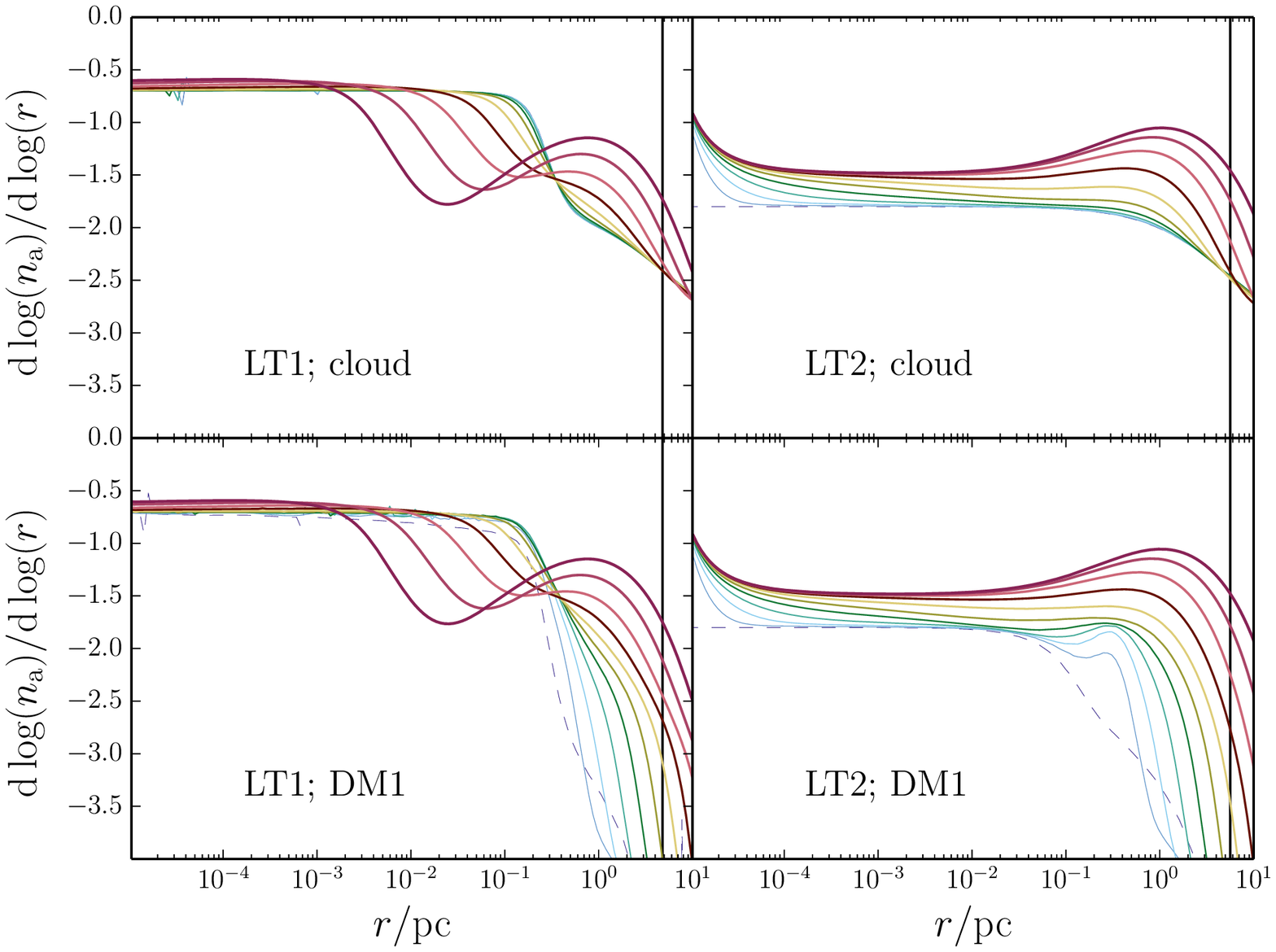}
\caption{\small Similar to Fig. \ref{fig:N_ast_r_F_lc_r_main_models}, here showing the planetesimal number density $n_\mathrm{a}(r,t)$ as a function of $r$ at the various times in the top four panels and the logarithmic number density derivative in the bottom four panels. }
\label{fig:n_ast_cer_main_models}
\end{figure}

\subsection{Results: distributions and disruption rates}
\label{sect:dyn_ff_ast:results}
The disruption rates at $t=10\,\mathrm{Gyr}$ are listed for various models in Table \ref{table:dis_rates}, assuming $N_\mathrm{a/\star}=2\times10^7$. In the first column the model number is listed; a succinct description of the models is given in Table \ref{table:dis_rates_description}. In model 1 the effect of scattering in energy and angular-momentum space by late-type stars is included; in other models late-type scattering is included, as well as other effects. For each of the models in Table \ref{table:dis_rates}, rates are quoted assuming the planetesimals were either formed in a large-scale cloud (`cloud') or in discs around stars (`disc'). In the latter case the stripping term, equation~(\ref{eq:F_strip_LT}), is included assuming either DM1 or DM2 (cf. Section~\ref{sect:models:disc_models}). Furthermore, rates are included for LT1 and LT2 (cf. Section~\ref{sect:models:GC:LT}). Below we discuss in more detail the results for the various models given in Table \ref{table:dis_rates}.

\subsubsection{Scattering by late-type stars only}
\label{sect:dyn_ff_ast:results:basic}
In the top four panels of Fig. \ref{fig:N_ast_r_F_lc_r_main_models} we show the the differential number of planetesimals, $N_\mathrm{a}(r,t)$, as a function of $r=\psi^{-1}(\Ec)$ at various times. Here, $N_\mathrm{a}(r,t)$ is defined as the number of planetesimals between radii $r$ and $r+\mathrm{d}r$, i.e. $N_\mathrm{a}(\psi(r),t) \, \mathrm{d} \psi(r) = N_\mathrm{a}(r,t) \, \mathrm{d}r$. The assumed model is model 1 (cf. Table \ref{table:dis_rates}). In the first (second) column results are shown for LT1 (LT2); the first row applies to formation in a cloud, whereas the second row applies to formation in discs (DM1). The radii of influence, defined via $M_\mathrm{LT}(r_h)=2M_\bullet$, where $M_\mathrm{LT}(r)$ is the distributed mass in late-type stars (cf. equation~\ref{eq:M_LT}), are indicated with the vertical black solid lines. 

In the top four panels of Fig. \ref{fig:n_ast_cer_main_models} the planetesimal number density $n_\mathrm{a}(r,t)$ is plotted at various times. It is computed from $f_\mathrm{a}(\Ec,t)$ using (e.g. \citealt[][3.2.1]{bookmerritt13})
\begin{align}
n_\mathrm{a}(r,t) = \int \mathrm{d}^3 v \, f_\mathrm{a}(v,t) = 4\pi \sqrt{2} \int_{-\infty}^{\psi(r)} \mathrm{d} \Ec \, f_\mathrm{a}(\Ec,t) \sqrt{\psi(r)-\Ec}.
\label{eq:n_from_f}
\end{align}
In Fig. \ref{fig:n_ast_cer_main_models} we also show in the bottom four panels the logarithmic derivative $\mathrm{d}\log(n_\mathrm{a})/\mathrm{d}\log(r)$. 

We show the initial distributions in Figs \ref{fig:N_ast_r_F_lc_r_main_models} and \ref{fig:n_ast_cer_main_models} with blue dashed lines. In the case of formation in a cloud the planetesimal number density $n_\mathrm{a}$ is proportional to the assumed number density of late-type stars at all radii. In the case of formation in discs this is the case at small radii because stripping is assumed to have occurred due to the tidal force of the SBH, whereas at larger radii planetesimals have not yet been stripped (cf. Section~\ref{sect:stripping:SBH}). Therefore the number density drops rapidly for radii $r\gtrsim 0.3 \, \mathrm{pc}$ (cf. the middle panel of Fig. \ref{fig:stripping_fraction_SBH}). 

\begin{figure}
\center
\includegraphics[scale = 0.435, trim = 0mm 0mm 0mm 0mm]{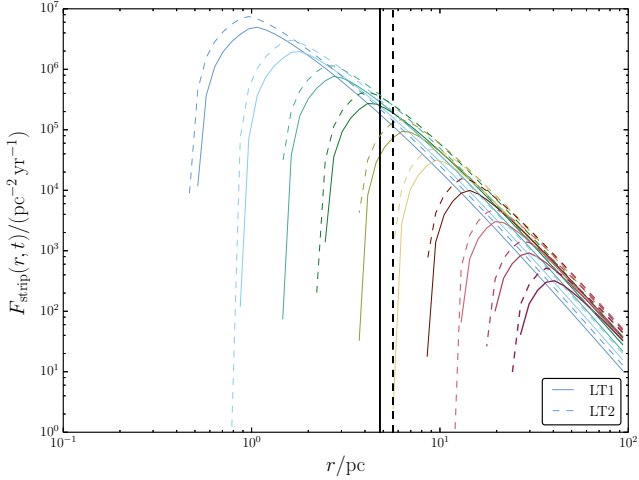}
\caption{\small The stripping flux $F_\mathrm{strip}(r,t)$ (cf. Section~\ref{sect:dyn_ff_ast:eq}) at the same times as in Fig. \ref{fig:N_ast_r_F_lc_r_main_models} assuming model 1 (cf. \ref{table:dis_rates_description}) and DM1. Solid lines: LT1; dashed lines: LT2. }
\label{fig:F_strip_r_main_models}
\end{figure}

\begin{figure}
\center
\includegraphics[scale = 0.435, trim = 0mm 0mm 0mm 0mm]{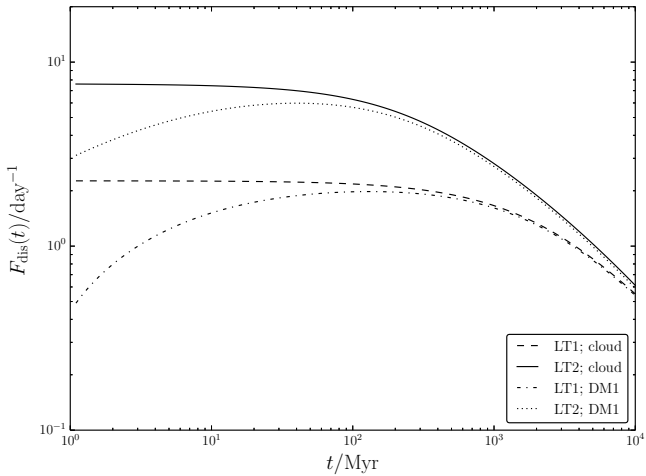}
\caption{\small The energy-integrated disruption rate $F_\mathrm{dis}(t)$ (cf. equation~\ref{eq:F_dis}) as a function of time assuming model 1 (cf. Table \ref{table:dis_rates_description}). Results are included for LT1 and LT2 and formation in a cloud and disc (DM1). }
\label{fig:F_lc_t_main_models}
\end{figure}

We show in Fig. \ref{fig:F_strip_r_main_models} the stripping flux $F_\mathrm{strip}(r,t)$ in the case of formation in discs at various times, for LT1 (solid lines) and LT2 (dashed lines). Here, we define $F_\mathrm{strip}(r,t)\,\mathrm{d}r$ as the number of stripped planetesimals per unit time between radii $r$ and $r+\mathrm{d}r$. Initially the stripping flux peaks at $r\approx 1\,\mathrm{pc}$. As time progresses planetesimals are stripped at progressively increasing radii. After 100 Myr the peak has shifted to outside the radius of influence, and the magnitude of the peak has decreased as well. There are no substantial differences in $F_\mathrm{strip}(r,t)$ between the distributions of late-type stars. This can be explained by the lack of influx of stripping by encounters at small radii $\lesssim r_b=0.3\,\mathrm{pc}$, at which the SBH is responsible for stripping, therefore the core in case of LT1 does not affect the stripping process. 

In the bottom panels of Fig. \ref{fig:N_ast_r_F_lc_r_main_models} the disruption flux $F_\mathrm{dis}(r,t)$ is shown, where $F_\mathrm{dis}(r,t)$ is the number of disrupted planetesimals per unit time between radii $r$ and $r+\mathrm{d}r$. For model LT1 this flux peaks at radii somewhat smaller than $r_h$, whereas for LT2 it is much less peaked. This difference can be attributed to the core in LT1. The flux at radii $>r_h$ is initially more than two orders of magnitude larger for formation in a cloud compared to formation in discs. However, as time progresses and planetesimals are stripped at increasing radii, this difference decreases. Despite this influx at large radii, the peak in the disruption flux does not shift to larger radii, but rather it becomes wider. 

In Fig. \ref{fig:F_lc_t_main_models} we show the energy-integrated disruption rates $F_\mathrm{dis}(t)$ (cf. equation~\ref{eq:F_dis}) for LT1 and LT2, and for both formation scenarios. Initially the rates for model LT2 are nearly an order of magnitude larger compared to model LT1. This can be attributed to the initially higher planetesimal density at small radii $r\lesssim 0.3\,\mathrm{pc}$ in case of LT2 because the late-type stars are assumed to be distributed in a cusp as opposed to a core (cf. the top four panels of Fig. \ref{fig:N_ast_r_F_lc_r_main_models}). While planetesimals are depleted at these radii the density drops, and as a consequence the loss cone flux decreases as well (cf. the bottom four panels of Fig. \ref{fig:N_ast_r_F_lc_r_main_models}). 

The stripping from stars causes the disruption rate to increase initially, whereas for formation in a cloud the rate only decreases with time. The differences are small, however: initially the rates in the former case are a factor of $\sim 2$ higher, and as time progresses the rates in the two cases gradually approach each other. After $\sim 100\,\mathrm{Myr}$ the rates become effectively indistinguishable, because stripping has progressed to beyond the radius of influence (cf. Fig. \ref{fig:F_strip_r_main_models}), where the disruption flux peaks (cf. the bottom four panels of Fig. \ref{fig:N_ast_r_F_lc_r_main_models}). By $t=10\,\mathrm{Gyr}$ the rates in all cases are nearly the same at $F_\mathrm{dis}(t=10\,\mathrm{Gyr})\approx 0.6\,\mathrm{day^{-1}}$ (cf. Table \ref{table:dis_rates}).

As mentioned in Section~\ref{sect:dyn_ff_ast:eq} the planetesimal steady-state distribution is expected to correspond to a number density $n(r)\propto r^{-3/2}$. The logarithmic slopes, shown in the bottom four panels of Fig. \ref{fig:n_ast_cer_main_models}, can be used to assess this. In the case of LT1 and by $t=10\,\mathrm{Gyr}$ the slopes are only consistent with $-3/2$ at a few radii $\ll r_h$, independent on the initial conditions. In particular, the slope is still close to the initial value of $-0.7$ (cf. Section~\ref{sect:models:GC:LT}) at radii $\lesssim 10^{-3} \, \mathrm{pc}$. Evidently, the core of late-type stars in the case of LT1 is unable to smoothen out the distribution at these radii, whereas this is the case in LT2 (note that the flattening in the later case for $r\lesssim 10^{-4} \, \mathrm{pc}$ arises from the boundary condition $f_\mathrm{a}(\Ec_\mathrm{lc},t)=0$). For both LT1 and LT2 there is a flattening in the number density near radii of $\sim 1 \, \mathrm{pc}$, causing a `bump' in the logarithmic slope. This can be attributed to the disruption flux, which peaks near this radius (cf. Fig. \ref{fig:N_ast_r_F_lc_r_main_models}).

\subsubsection{Stripping from early-type stars}
\label{sect:dyn_ff_ast:results:ET_strip}
As mentioned in Section~\ref{sect:models:GC:ET} planetesimals in the GC could also form in debris discs around early-type stars. This source of planetesimals has been included in model 2 (cf. Table \ref{table:dis_rates}) by adding the term $F_\mathrm{strip;ET}(\Ec)$ (cf. equation~\ref{eq:F_strip_ET}) to the right-hand side of equation~(\ref{eq:FP_main}). We find that the inclusion of this term has virtually no effect on the planetesimal distribution at all times. This is also reflected by the disruption rates, which change by less than one per cent (cf. Table \ref{table:dis_rates}). Considering that equation~(\ref{eq:F_strip_ET}) is likely an upper limit for the supply of planetesimals from early-type stars, we conclude that stripping from early-type stars can be safely neglected compared to stripping from the much more numerous late-type stars.

\subsubsection{Scattering by a cusp of stellar black holes}
\label{sect:dyn_ff_ast:results:bh_scat}
In models 3 and 4 scattering in energy and angular momentum is taken into account assuming a (hypothesized) cusp of stellar black holes close to the SBH. In the case of model 4 the black hole number density is time-dependent and is assumed to increase linearly with time, reaching the number density that we assume at all times in model 3 (cf. Section~\ref{sect:models:GC:stellar_black_holes}), at $t=10\,\mathrm{Gyr}$. 

\begin{figure}
\center
\includegraphics[scale = 0.435, trim = 0mm 0mm 0mm 0mm]{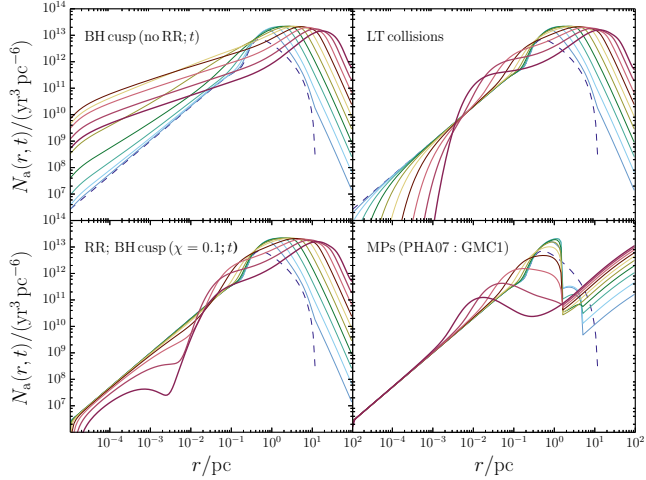}
\caption{The quantity $N_\mathrm{a}(r,t)$ for four models (models 4, 5, 14 and 15, cf. Table \ref{table:dis_rates_description}) as a function of $r$ at various times. In all panels LT1 and formation in a disc (DM1) are assumed. }
\label{fig:N_ast_r_comparison}
\end{figure}

\begin{figure}
\center
\includegraphics[scale = 0.435, trim = 0mm 0mm 0mm 0mm]{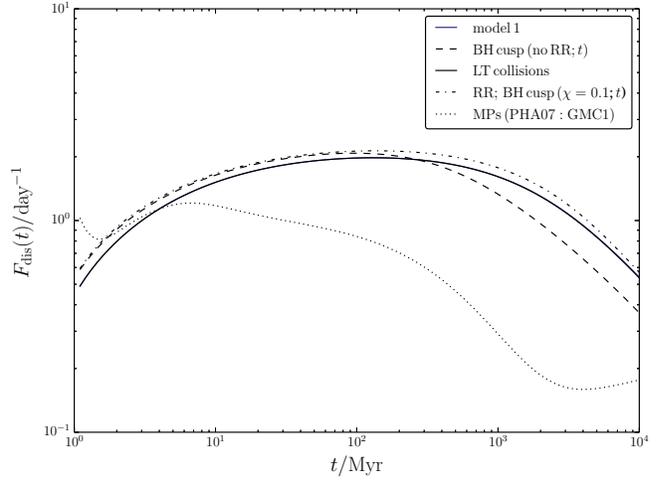}
\caption{The total disruption rate $F_\mathrm{dis}(t)$ as a function of time (cf. equation~\ref{eq:F_dis}) for model 1 and the models included in Fig. \ref{fig:N_ast_r_comparison}. In all cases LT1 and DM1 are assumed.}
\label{fig:F_lc_t_comparison}
\end{figure}

We show in the top left panel of Fig. \ref{fig:N_ast_r_comparison} the quantity $N_\mathrm{a}(r,t)$ for the case of a time-dependent black hole cusp and for LT1 and DM1. The effect of a cusp of stellar black holes is initially to increase the density of planetesimals at small radii, $r\lesssim 0.1 \, \mathrm{pc}$. At later times this density decreases again. This is reflected in the disruption rate, which is shown as a function of time in Fig. \ref{fig:F_lc_t_comparison} (black dashed line). Initially the rate is slightly higher compared to the model 1 (blue solid line), but after $\approx 200 \, \mathrm{Myr}$ it drops slightly below the latter rate.

\subsubsection{Collisions}
\label{sect:dyn_ff_ast:results:col}[][5.5.1]{bookmerritt13}
We consider three different models in which the effects of physical collisions are taken into account. In model 5 collisions between late-type stars and planetesimals are considered; the initial stellar mass is assumed to be $1\,\mathrm{M}_\odot$, and the mass and radius at subsequent times are adopted from a stellar evolution model with metallicity $Z=0.02$ (the \textsc{SSE} code, \citealt{hpt00}, is used as implemented in \textsc{AMUSE}, \citealt{pelupessy_ea13,pmepv13}). In the latter model the radius changes only little during 10 Gyr; we have also included a model, model 6, in which the stellar mass is assumed to be $2\,\mathrm{M}_\odot$ (resulting in faster evolution). In the third model, model 7, planetesimal-planetesimal collisions are taken into account with a number of simplifying assumptions (cf. Section~\ref{app:FP_main:col}). 

The quantity $N_\mathrm{a}(r,t)$ for model 5 is plotted in the the top right panel of Fig. \ref{fig:N_ast_r_comparison} (results for models 6 and 7 are similar). It is reduced substantially compared to model 1 only at radii $\lesssim 10^{-2} \, \mathrm{pc}$ (cf. the top right panel of Fig. \ref{fig:N_ast_r_comparison}). The disruption flux peaks at much larger radii (cf. Fig. \ref{fig:N_ast_r_F_lc_r_main_models}). Therefore the effect of collisions on the disruption rates are very small. This is illustrated in Fig. \ref{fig:F_lc_t_comparison}, in which the rates according to model 5 (black solid line) are indistinguishable, from an observational perspective, from the rates in the model 1 (blue solid line).

\subsubsection{Resonant relaxation}
\label{sect:dyn_ff_ast:results:RR}
In six models, models 8-13, the effects of RR are included (cf. Section~\ref{app:FP_main:RR}). In models 8 and 9 RR is assumed to arise from the late-type stars; in models 10-13 RR is assumed to arise from a cusp of stellar black holes, with and without a time dependence (cf. Section~\ref{sect:models:GC:stellar_black_holes}). Two values of the efficiency of RR, $\chi_\mathrm{RR}$ (cf. Section~\ref{app:FP_main:RR}), are assumed: $\chi_\mathrm{RR}=1$ and $\chi_\mathrm{RR} = 0.1$. 

In the case of RR arising from late-type stars, $N_\mathrm{a}(r,t)$ is not affected noticeably compared to the case without RR. The disruption rates are enhanced only a few per cent, even if $\chi_\mathrm{RR}=1$ (cf. Table \ref{table:dis_rates}). A cusp of stellar black holes affects $N_\mathrm{a}(r,t)$ more strongly, but this is sensitive to the RR efficiency $\chi_\mathrm{RR}$ and whether or not the cusp is assumed to be time-dependent. 

In the bottom left panel of Fig. \ref{fig:N_ast_r_comparison} $N_\mathrm{a}(r,t)$ is shown as a function of $r$ for various times assuming a time-dependent cusp of stellar black holes with $\chi_\mathrm{RR}=0.1$ (model 13). This model can be considered as the most realistic model of RR with a black hole cusp. Only at late times, $t\sim 10 \, \mathrm{Gyr}$, and at small radii, $r\lesssim 10^{-2}\,\mathrm{pc}$, is RR effective at reducing $N_\mathrm{a}(r,t)$. Nevertheless, the energy-integrated disruption rates at $t=10\,\mathrm{Gyr}$ are increased by at most 3 per cent compared to model 1 (cf. Table \ref{table:dis_rates}).

\subsubsection{Massive perturbers}
\label{sect:dyn_ff_ast:results:MP}
In model 14 the effect of massive perturbers is taken into account (cf. Section~\ref{sect:models:GC:MP}). Contrary to what might be expected, the disruption rates at $t=10\,\mathrm{Gyr}$ are a factor of $\sim 2$ lower compared to those of model 1. The massive perturbers considered here are located at radii $>1.5\,\mathrm{pc}$ (cf. table 2 of PHA07), which corresponds to the full loss cone regime in the case of relaxation by late-type stars only (cf. the black vertical dotted lines in the bottom four panels of Fig. \ref{fig:N_ast_r_F_lc_r_main_models}). The full loss cone rate corresponds to the maximum loss rate, and this explains why the disruption rates are initially only modestly larger, by a factor of $\sim 2$, as shown in Fig. \ref{fig:F_lc_t_comparison} (dotted lines). Relaxation in energy is also assumed to be much faster, however, therefore at the radial range of the massive perturbers, $1.5\,\mathrm{pc}<r<100\,\mathrm{pc}$, planetesimals are efficiently transported from radii $r\sim 1\, \mathrm{pc}$ to $r\gg 1\,\mathrm{pc}$, as illustrated in the bottom right panel of Fig. \ref{fig:N_ast_r_comparison}. Because the loss cone flux still peaks near $1\,\mathrm{pc}$ (cf. Fig. \ref{fig:N_ast_r_F_lc_r_main_models}), this implies that the disruption rate must drop rapidly, which is reflected in Fig. \ref{fig:F_lc_t_comparison}. Nevertheless, after 10 Gyr the disruption rate is still $\approx 0.2 \, \mathrm{day^{-1}}$, which is only a factor of $\approx 3$ lower compared model 1. 

\subsubsection{Semi-analytic solutions}
\label{sect:dyn_ff_ast:results:semi_an}
Equation~(\ref{eq:FP_main}) was solved numerically. We also obtained approximate semi-analytic solutions to this equation for model 1. In order to solve the equation, we assume that a steady-state in energy applies at all times. With this assumption, and setting the terms $F_\mathrm{col}$ and $F_\mathrm{RR}$ to zero in accordance with model 1, we find semi-analytical solutions of the disruption flux as a function of time (refer to Appendix \ref{app:FP_semi_an} for details). 

\begin{figure}
\center
\includegraphics[scale = 0.435, trim = 0mm 0mm 0mm 0mm]{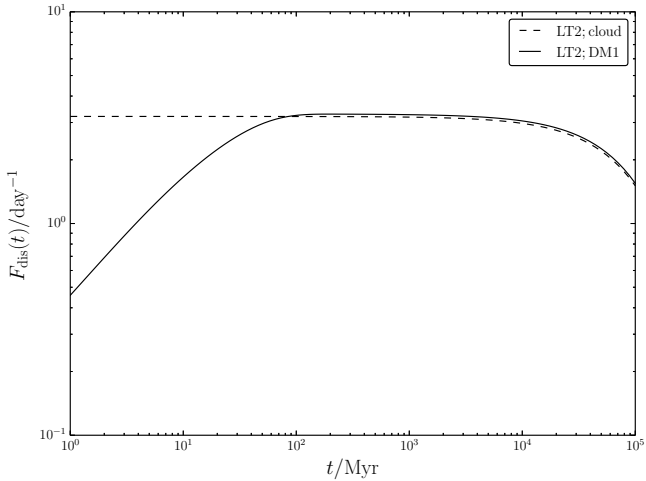}
\caption{\small The energy-integrated disruption rate $F_\mathrm{dis}(t)$ as a function of time as in Fig. \ref{fig:F_lc_t_main_models}, here according to approximate semi-analytic solutions to equation~(\ref{eq:FP_main}), for which it is assumed that a steady-state in energy applies at all times (cf. Section~\ref{sect:dyn_ff_ast:results:semi_an}). Rates are shown for LT2 and the two formation scenarios. }
\label{fig:F_lc_t_semi_an}
\end{figure}

These solutions are shown in Fig. \ref{fig:F_lc_t_semi_an} for LT2 and the two formation scenarios. The time-scale of the decrease of the disruption rate after $\sim 100\, \mathrm{Myr}$, $\gg 10^4 \, \mathrm{Myr}$, is too long compared to the numerical solutions (cf. Fig. \ref{fig:F_lc_t_main_models}). Nevertheless, the semi-analytic solutions yield disruption rates of the same order of magnitude, therefore they should be adequate for the purposes of scaling the rates to different galactic nuclei (cf. Section~\ref{discussion:scaling}). 

\section{Discussion}
\label{sect:discussion}
\subsection{Comparison to observations: constraints on $N_\mathrm{a/\star}$}
\label{sect:discussion:constraints}
The disruption rates at $t=10\,\mathrm{Gyr}$ as shown in Table \ref{table:dis_rates} are generally $\approx 0.6 \, \mathrm{day^{-1}}$ for models 1-13; in the case of model 14 (massive perturbers) the rate is slightly lower, $\approx 0.2 \, \mathrm{day^{-1}}$. These rates are robust in the sense that there is no strong dependence on the assumed distribution of late-type stars (LT1 vs. LT2), nor on the formation scenario (cloud vs. disc). In particular, we note that our rates vary only a few per cent between the two disc models DM1 and DM2. This is because stripping by gravitational encounters occurs early in the evolution, and it is very effective at stripping all planetesimals at radii $\lesssim r_h$ for both DM1 and DM2, even though in DM2 planetesimals are more tightly bound to the star (cf. the bottom panel of Fig. \ref{fig:stripping_fraction_enc}). 

Based on 3 Ms {\it Chandra} observations in 2012 the observed X-ray flaring rate is $1.1^{+0.2}_{-0.1} \, \mathrm{day^{-1}}$ \citep{neilsen_ea13}. Strictly speaking, this would seem inconsistent with our rate of $\approx 0.6 \, \mathrm{day^{-1}}$ ($5\sigma$ deviation). 

We emphasize, however, that in Section~\ref{sect:dyn_ff_ast} we assumed $N_\mathrm{a/\star}=2\times 10^7$ and that this quantity is poorly constrained (the main uncertainty in $N_\mathrm{a/\star}$ is the total mass of planetesimals per star). Our result can therefore be used to constrain $N_\mathrm{a/\star}$ within the framework of our other assumptions. Using that the disruption rate is linearly proportional to $N_\mathrm{a/\star}$ (cf. Section~\,\ref{discussion:scaling}), it directly follows that according to models 1-13, $N_\mathrm{a/\star} \approx 3.7 \times 10^7$, and according to model 14, $N_\mathrm{a/\star}\approx 1.1 \times 10^8$. 

\subsection{Internal scattering of planetesimals by planets}
\label{sect:discussion:internal_scat}

In \S\,\ref{sect:dyn_ff_ast}, internal processes in the debris discs were not considered. Likely the most important of these processes is scattering of planetesimals by planets bound to the star, which could lead to the ejection of a significant fraction of planetesimals from the star, analogously to the Nice model \citep{gomes_ea_05}. The precise fraction of ejected planetesimals likely depends strongly on the distribution of planets around stars in the GC, which is currently completely unconstrained.  

The scenarios considered in \S\,\ref{sect:dyn_ff_ast} can be interpreted as two extreme cases of internal scattering. In our cloud scenario, the number of planetesimals was chosen to be consistent with the number of stars and the number of planetesimals per star, and the planetesimals were assumed to be formed with an orbital distribution around the SBH similar to that of the stars. Modulo a likely delay between formation and ejection, this is consistent with an extreme case of internal scattering where {\it all} planetesimals are ejected from the debris disc. In contrast, in our disc scenario ejection was only assumed to occur due to encounters with other stars, i.e. no internal processes are taken into account. 

The above implies that our cloud and disc scenarios can be used to evaluate the effect of internal processes on the flaring rate. The weak dependence of the latter rate on the assumed scenario, in particular at late times (cf. Figure \ref{fig:F_lc_t_main_models}), suggests that this effect is very weak. 

\subsection{Scaling of the disruption rate: tidal disruption of planets}
\label{discussion:scaling}
In Section~\ref{sect:dyn_ff_ast} the number of planetesimals per star was assumed to be $N_\mathrm{a/\star}=2\times 10^7$, consistent with observations of debris discs around stars in the Solar neighbourhood (ZNM12; \citealt{wyatt08}). Other important parameters that were assumed are the masses $M_\bullet=4\times 10^6\,\mathrm{M}_\odot$ and $m_\mathrm{LT}=1\,\mathrm{M}_\odot$, and the distributed mass in late-type stars within $r_0=1\,\mathrm{pc}$, $M_0=1.5\times 10^6 \, \mathrm{M}_\odot$ (cf. Section~\ref{sect:models:GC:LT}), or, equivalently, the number of late-type stars within $r_0$, $N_0=M_0/m_\mathrm{LT} = 1.5 \times 10^6$. 

The SBHs with masses $\sim 10^6 \, \mathrm{M}_\odot$ in several nearby spiral galaxies have quiescent X-ray luminosities of $\sim 10^{37}-10^{39}$ erg/s (\citealt{baganoff_ea03} and references therein). These luminosities are several orders of magnitude larger than the typical X-ray flare luminosity associated with the tidal disruption of planetesimals, $\sim10^{34}-10^{35}$ erg/s, making it unlikely that such flares could be observed in galactic nuclei other than the GC. 

This conclusion could be different for the tidal disruption of planets by the SBH, which would produce a much more luminous flare. For example, ZNM12 estimated that the tidal disruption of a Jupiter-mass gas giant would produce a flare with an X-ray luminosity of $\sim 2\times 10^{41}$ erg/s, and this would therefore dominate the quiescent luminosity. Such a flare could be observed with e.g. the {\it INTEGRAL} space telescope up to a distance of $\sim 50 \, \mathrm{Mpc}$; the number of galaxies within this distance is $\sim5000$ \citep{nw13}. 

For this purpose we obtain an approximate scaling of the disruption rate based on the semi-analytic solutions to equation~(\ref{eq:FP_main}) that were discussed in Section~\ref{sect:dyn_ff_ast:results:semi_an}, with the inclusion of the stripping term (i.e. assuming formation in discs) and the angular-momentum loss cone term arising from scattering with late-type stars. As in Section~\ref{sect:dyn_ff_ast:results:semi_an} we assume that at all times a steady-state in the planetesimal distribution function has been reached with respect to the orbital energies. Furthermore, in order to find an analytic scaling we approximate the effect of stripping by gravitational encounters as an instantaneous process, we assume a single power-law number density of the perturbing stars, $n_\star\propto r^{-\gamma}$, and we neglect the stellar potential. 

With these assumptions and approximations we find the following order-of-magnitude estimate of the disruption rate (refer to Appendix \ref{app:scaling} for more details) 
\begin{align}
F_\mathrm{dis}(t) \sim \frac{N_\mathrm{a/\star} N_0}{\tau} \exp(-t/\tau) \left [ 1 + \exp(t_c/\tau) \right ],
\label{eq:F_lc_scaling_t}
\end{align}
where $\tau$ is given by
\begin{align}
\nonumber \tau^{-1} & \equiv \left [ \int \mathrm{d} \Ec \, S_\mathrm{lc}(\Ec) \right ] \left [ 4 \pi^2 \int \mathrm{d} \Ec \, p(\Ec) \right ]^{-1} \\
\nonumber & \sim N_0 (Gm_\star)^2 (GM_\bullet r_0)^{-3/2} \log(\Lambda) \left ( \frac{2M_\bullet}{m_\star} N_0^{-1} \right )^{\frac{\gamma-3/2}{\gamma-3}}. \\
\label{eq:F_lc_scaling}
\end{align}
Here, we neglected factors of order unity. The quantity $t_c$ is the time-scale for which most planetesimals are assumed to have been stripped by gravitational encounters; $t_c \sim 100\, \mathrm{Myr}$ for the GC (cf. Section~\ref{sect:stripping:enc}). Equations~(\ref{eq:F_lc_scaling_t}) and (\ref{eq:F_lc_scaling}) give for the GC parameters and $\gamma=2$ at $t=10\,\mathrm{Gyr}$
\begin{align}
\nonumber & F_\mathrm{lc}(t=10\,\mathrm{Gyr}) \approx 6 \, \mathrm{day^{-1}} \left ( \frac{N_\mathrm{a/\star}}{2\times 10^7} \right ) \left ( \frac{m_\star}{\mathrm{M}_\odot} \right )^{5/2} \\
\nonumber &\quad \times \left (\frac{N_0}{1.5\times 10^6} \right )^{5/2} \left ( \frac{r_0}{1.0 \, \mathrm{pc}} \right )^{-3/2} \left ( \frac{M_\bullet}{4\times 10^6 \, \mathrm{M}_\odot} \right )^{-2} \\
&\quad \times \log \left ( \frac{M_\bullet}{4\times 10^6 \, \mathrm{M}_\odot} \frac{1\,\mathrm{M}_\odot}{2m_\star} \right ). 
\label{eq:F_lc_scaling_GC}
\end{align}
This estimate is correct within an order of magnitude (cf. Table \ref{table:dis_rates}). The scaling is linear with $N_\mathrm{a/\star}$, which is intuitively easy to understand. The scaling with $N_0$ is $F_\mathrm{lc}\propto N_0^{5/2}$. The latter can be understood as follows: increasing the number of stars increases the total supply of planetesimals but also accelerates the rate of scattering into the loss cone, hence the dependence is stronger than linear.

We use the above scaling relation to estimate the frequency at which {\it INTEGRAL} could detect planet disruptions. In equation~(\ref{eq:F_lc_scaling}) we set $r_0=r_h$ and $N_0 = 2M_\bullet/m_\star$ with $m_\star=1\,\mathrm{M}_\odot$, i.e. corresponding to the sphere of influence (with this choice there is no dependence of $\tau$ on $\gamma$). The radius of influence $r_h$ is subsequently calculated from $r_h = GM_\bullet/\sigma^2$, where the velocity dispersion $\sigma$ is computed from the $M_\bullet-\sigma$-relation \citep{ferrarese_merritt00}. With these assumptions, the resulting expression for the disruption rate is a function of SBH mass, $F_\mathrm{dis} = F_\mathrm{dis}(M_\bullet)$. Here, we set $F_\mathrm{dis}(M_\bullet) = 0$ for $M_\bullet\gtrsim10^7 \, \mathrm{M}_\odot$ because for the latter masses the relaxation time-scale is longer than a Hubble time, and steady-state in energy can no longer be assumed (e.g. \citealt[][Eq. 3.6]{bookmerritt13}). (In addition, for $M_\bullet \gtrsim 10^8 \, \mathrm{M}_\odot$ the tidal disruption radius of the planet, similar to that of a $1\,\mathrm{M}_\odot$ star, is smaller than the Schwarzschild radius (e.g. Fig. 6.1 of \citealt{bookmerritt13}), implying that the planet would not be disrupted and eventually produced a potentially observable flare, but captured whole. )

Subsequently, we average $F_\mathrm{dis}$ over the observed mass function of SBHs in the centres of local galaxies,
\begin{align}
\overline{F}_\mathrm{dis} = \frac{\int_{M_{\bullet,\mathrm{low}}}^{M_{\bullet,\mathrm{up}}} \mathrm{d} M_\bullet \, F_\mathrm{dis}(M_\bullet) \, \mathrm{d}N/\mathrm{d}M_\bullet}{\int_{M_{\bullet,\mathrm{low}}}^{M_{\bullet,\mathrm{up}}} \mathrm{d} M_\bullet \, \mathrm{d}N/\mathrm{d}M_\bullet},
\label{eq:F_planet_av}
\end{align}
where $\mathrm{d}N/\mathrm{d}M_\bullet$ is given by Eq. (3) and Table 2 (first row) of \citet{vika_ea09} and $M_{\bullet,\mathrm{low}} = 10^6\,\mathrm{M}_\odot$ and $M_{\bullet,\mathrm{up}} = 10^{10}\,\mathrm{M}_\odot$. 

Multiplying the average rate implied by equation~(\ref{eq:F_planet_av}) by the number of galaxies within 50 Mpc, $\sim 5000$, we find the following estimate for the rate of planet disruptions that are observable by {\it INTEGRAL},
\begin{align}
F_\mathrm{dis,planet} \sim 0.05 N_\mathrm{p/\star} \, \mathrm{yr^{-1}},
\label{eq:F_planet}
\end{align}
where $N_\mathrm{p/\star}$ is the number of planets per star. We therefore expect an observable planet disruption roughly every decade.

\subsection{Changes of the stellar orbit prior to stripping}
\label{sect:discussion:orbit_change}
The debris disc of a star that was formed with pericentre distance $r_p \gg r_\mathrm{H} \equiv d_2 (3M_\bullet/m_\star)^{1/3}$ from the SBH is not tidally stripped by the SBH, where $d_2$ is the outer radius of the debris disc (cf. Section~\ref{sect:stripping:SBH}). Gravitational encounters will gradually strip the debris disc over time at a time-scale on the order of $t_\mathrm{strip}\sim 100 \,\mathrm{Myr}$ in the GC (cf. Section~\ref{sect:stripping:enc}). In some cases, however, the time-scale for gravitational encounters with other stars to change the orbital properties of the {\it parent star} can be shorter than $t_\mathrm{strip}$. 

The time-scale for the orbital energy to change by order itself due to non-resonant relaxation (NRR), $t_\Ec(\Ec) \equiv [\langle (\Delta \Ec)^2 \rangle /\Ec^2 ]^{-1}$, where $\langle (\Delta \Ec)^2 \rangle$ is the second-order energy diffusion coefficient, is typically longer than the stripping time-scale. To illustrate this we show the ratio $t_\mathrm{strip}(\Ec)/t_\mathrm{\Ec}(\Ec)$ as a function of $r=\psi^{-1}(\Ec)$ in the top panel of Fig. \ref{fig:special_loss_cone_orbits}. Here, $t_\mathrm{strip}$ is the orbit-averaged stripping time-scale where $a_\mathrm{bin}=d_2$ (cf. Section~\ref{sect:stripping:enc}). Furthermore, $t_\mathrm{\Ec}(\Ec)$ is computed from the late-type distribution function using the expression for $\langle (\Delta \Ec)^2 \rangle$ in Eq. (21) of \citet{ck78}, without neglect of the stellar potential. At most radii $t_\mathrm{strip}(\Ec)/t_\mathrm{\Ec}(\Ec) \ll 1$, demonstrating that stripping occurs faster than energy relaxation. However, the time-scale $t_L \sim (L/L_\mathrm{c})^2 \,t_\Ec$ for encounters to change the {\it orbital angular momentum} by order itself, where $L_\mathrm{c}$ is the angular momentum of a circular orbit, can be much shorter if the orbit is highly eccentric. 

Nevertheless, even if $t_L \ll t_\mathrm{strip}$ we do not expect our results in Section~\ref{sect:dyn_ff_ast} to be much affected because NRR does not affect the {\it statistical} properties of the orbits of the parent stars in angular momentum. This is because the steady-state angular-momentum distribution due to NRR is an isotropic distribution (i.e. $f(\Ec,L) = f(\Ec)$), which is the same distribution that was assumed in equation~(\ref{eq:FP_main}) (an exception is that $f(\Ec,L) \propto \log(L)$ for orbits very close to the SBH, \citealt{ck78}).

\subsection{Special case: a burst of flares?}
\label{sect:discussion:burst}
In Sections~\ref{sect:stripping} and \ref{sect:dyn_ff_ast} it was assumed that the planetesimals resulting in flares are bound to the SBH prior to their disruption, either because they were formed in a large-scale cloud bound to the SBH, or they were born in discs and stripped from their parent star. Here, we consider a special case in which the planetesimals are bound to a star prior to being disrupted by the SBH, possibly resulting in a burst of flares.

\begin{figure}
\center
\includegraphics[scale = 0.435, trim = 0mm 0mm 0mm 0mm]{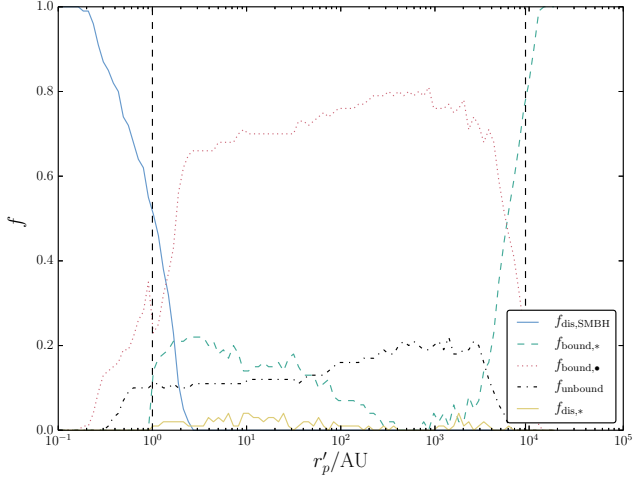}
\caption{\small Fractions of the various outcomes as a function of the pericentre distance $r_p'$ in the $N$-body simulations of a star with a disc approaching the SBH on a highly eccentric orbit, discussed in Section~\ref{sect:discussion:burst}. Blue solid line: planetesimals that are disrupted by the SBH (i.e. pass the SBH within 1 AU); green dashed line: planetesimals that remain bound to the star; red dotted line: planetesimals that become bound to the SBH; black dot-dashed line: planetesimals that formally become unbound from the SBH (with low escape velocities, cf. Section~\ref{sect:discussion:burst}); solid yellow line: planetesimals that collide with the star (the stellar radius is assumed to be $1\,\mathrm{R}_\odot$). The left (right) black vertical dashed lines indicate $r_\mathrm{lc}$ ($r_\mathrm{H}$).}
\label{fig:fractions_special_lc}
\end{figure}

\begin{figure}
\center
\includegraphics[scale = 0.435, trim = 0mm 0mm 0mm 0mm]{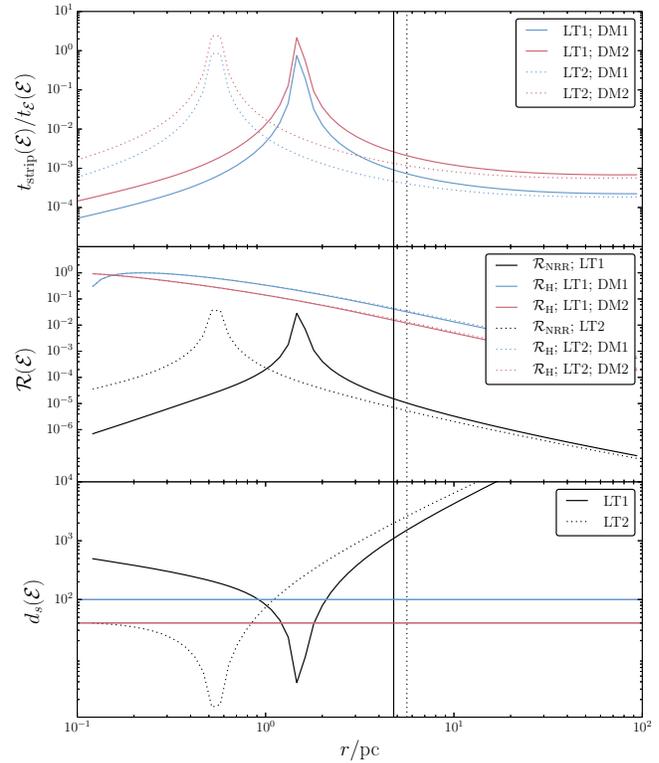}
\caption{\small Various quantities as a function of $r=\psi^{-1}(\Ec)$ discussed in Section~\ref{sect:discussion:burst}. Two distributions of late-type stars are assumed: LT1 and LT2 (cf. Section~\ref{sect:models:GC:LT}). Blue (red) lines apply to DM1 (DM2) (cf. Section~\ref{sect:models:disc_models}). {\bf Top panel}: the ratio of the orbit-averaged stripping time-scale to the energy time-scale. {\bf Middle panel}: the value of $R$, $R_\mathrm{NRR}$, required for a gravitational encounter to sufficiently decrease the pericentre distance to produce a burst of flares, compared to $R_\mathrm{H}$, the value of $R$ for which the SBH strips the disc and a burst would not occur. 
{\bf Bottom panel}: the stripping distance $d_s$ at which the strong encounter would strip part or all of the disc. }
\label{fig:special_loss_cone_orbits}
\end{figure}

This special case could arise if the orbit of the star is highly eccentric with pericentre distance close to but slightly larger than the Hill radius $r_\mathrm{H}$ (cf. Section~\ref{sect:discussion:orbit_change}), and relaxation into an even more eccentric orbit subsequently drives the pericentre distance to $r_p'<r_\mathrm{H}$. The tidal disruption radius of the star, $r_\mathrm{TD,\star} \approx R_\star \, (M_\bullet/m_\star)^{1/3}\approx 0.7 \, \mathrm{AU}$ (assuming $m_\star=1\,\mathrm{M}_\odot$, $R_\star=1\,\mathrm{R}_\odot$ and $M_\bullet=4\times 10^6\,\mathrm{M}_\odot$), is approximately equal to that of the planetesimals. Therefore, if the new pericentre distance is $r_p' \approx r_\mathrm{lc}=1\,\mathrm{AU}$ then we expect the star to be tidally disrupted by the SBH; the resulting flare would outshine the flares resulting from tidally disrupted planetesimals by many orders of magnitude. 

However, if $r_p'$ is slightly larger than the tidal disruption radius of the star then, owing to the extended size of the debris disc, some or all of the planetesimals might be stripped from the star, pass within $1 \,\mathrm{AU}$ of the SBH and ultimately produce a flare, whereas the star is not tidally disrupted. To investigate this case, we performed a series of $N$-body simulations of a star with mass $1\,\mathrm{M}_\odot$ with a debris disc approaching the SBH ($M_\bullet=4\times 10^6 \, \mathrm{M}_\odot$) on a highly eccentric orbit. These simulations, similar to those of Section~\ref{sect:stripping:SBH}, were carried out with the \textsc{hermite0} code \citep{hmm95} in the \textsc{AMUSE} framework \citep{pelupessy_ea13,pmepv13}. The orbit of the star has a semimajor of 1 pc, and the pericentre distance $r_p'$ is varied between $0.1\,r_\mathrm{lc}$ and $2\,r_\mathrm{H}$, where $r_\mathrm{lc}=1\,\mathrm{AU}$ and $r_\mathrm{H} \equiv d_2 (3M_\bullet/m_\star)^{1/3} \approx 9.2\times 10^3 \, \mathrm{AU}$. The debris disc is sampled consistently with DM1 (cf. Section~\ref{sect:models:disc_models}) and the initial distance of the star to the SBH is $4\,r_\mathrm{H}$. 

In Fig. \ref{fig:fractions_special_lc} the fractions of the various outcomes in these simulations are shown as a function of $r_p'$. As expected, for $r_p'\gtrsim r_\mathrm{H}$ (indicated with the right black vertical dashed line) all planetesimals remain bound to the star. For $r_\mathrm{lc} < r_p' < r_\mathrm{H}$ most ($\sim 0.8$) planetesimals are stripped and become bound to the SBH. A smaller fraction ($\sim 0.2$) becomes formally unbound from the SBH. We note that the escape speeds of these unbound planetesimals are low; the typical speed is $50 \, \mathrm{km/s}$. Therefore it is likely that subsequent gravitational perturbations from stars could cause these planetesimals to be again bound to the SBH. 

For $r_p' \lesssim r_\mathrm{lc}$ (indicated with the left black vertical dashed line) most planetesimals are disrupted by the SBH. Disruption of planetesimals by the SBH also occurs for $r_p>r_\mathrm{lc}$, but only for $r_p' \lesssim 2 \, \mathrm{AU}$. Because the latter pericentre distance is close to $r_\mathrm{TD,\star}$, we expect that in this case the star would be either partially or fully disrupted, and therefore the resulting flare would be completely dominated by the star. 

A burst of flares originating from multiple planetesimals can also be excluded based on an examination of the likelihood that a star with a debris disc would receive the perturbation necessary to change the pericentre from $r_p\geq r_\mathrm{H}$ to $r_p'\ll r_\mathrm{H}$. The following conditions need to be satisfied\footnote{The basic problem discussed here is similar to that of \citet{ml12}.}:
\begin{enumerate}
\item The initial pericentre distance $r_p = \theta \, r_\mathrm{H}$ with $\theta \geq 1$, whereas the final pericentre distance $r_p' = \epsilon \, r_\mathrm{H}$ with $r_\mathrm{lc}/r_\mathrm{H} < \epsilon \ll 1$. If $r_p < r_\mathrm{H}$, then the star would have been stripped by the SBH at an earlier time. If $\epsilon$ is smaller than unity but not small enough, then most planetesimals will be stripped by the SBH but not pass within $r_\mathrm{lc}=1\,\mathrm{AU}$ of the SBH (cf. Fig. \ref{fig:fractions_special_lc}). This also implies that the gravitational perturbation causing the pericentre change must act over a (radial) orbital time-scale $P_r(\Ec)$. 
\item The gravitational perturbations must be strong enough to produce the required decrease of the pericentre distance, but may not be too strong to disrupt the debris disc. 
\end{enumerate}

We assume that the pericentre changing encounters occur at apocentre and instantaneously change the apocentre velocity, whereas the apocentre distance and orbital energy remain unaffected. The change of the pericentre distance is therefore given by
\begin{align}
\Delta r_p \approx \frac{\partial r_p(\Ec,\Rc)}{\partial \Rc} \, \Delta \Rc,
\label{eq:r_p_change}
\end{align}
where $\Rc\equiv [L/L_\mathrm{c}(\Ec)]^2$ is a normalized angular-momentum variable. Furthermore, during $P_r(\Ec)$ the change of angular momentum due to NRR is assumed to occur through a random walk,
\begin{align}
\Delta L = L_\mathrm{c}(\Ec) \left ( \frac{P_r(\Ec)}{t_\Ec(\Ec)} \right )^{1/2}.
\label{eq:random_walk}
\end{align}

Combining equations~(\ref{eq:r_p_change}) and (\ref{eq:random_walk}) and writing the pericentre change as $\Delta r_p/r_p = \epsilon/\theta-1\approx-1$ we find the following relation for the angular momentum satisfying the requirement of the pericentre change,
\begin{align}
\Rc_\mathrm{NRR}(\Ec) \approx \frac{4P_r(\Ec)}{t_\Ec(\Ec)} \left ( \frac{\epsilon}{\theta} - 1 \right )^{-2} \left ( \frac{\partial \log r_p(\Ec,\Rc)}{\partial \log \Rc} \right )^2.
\label{eq:R_req_enc}
\end{align}
The logarithmic derivative factor in equation~(\ref{eq:R_req_enc}) is a function of $\Rc$, implying that equation~(\ref{eq:R_req_enc}) cannot in general be solved analytically for $\Rc_\mathrm{NRR}$. However, this factor is a weak function of $\Rc$ for $\Rc\ll1$ and close to unity in this limit (in the case when the stellar potential is neglected, it is exactly equal to unity); for the computations below we set it equal to unity. 

The quantity $\Rc_\mathrm{NRR}(\Ec)$ is shown as a function of $r=\psi^{-1}(\Ec)$ with black solid (dotted) lines in the middle panel of Fig. \ref{fig:special_loss_cone_orbits} for late-type distribution LT1 (LT2), assuming $\epsilon/\theta = 10^{-3}$. There is a peak where the energy relaxation time-scale $t_\mathrm{\Ec}(\Ec)$ is shortest. In addition we show in the second panel of Fig. \ref{fig:special_loss_cone_orbits} with the blue (red) lines the quantity $\Rc_\mathrm{H}$, defined as the value of $R$ corresponding to $r_p=r_\mathrm{H}$ assuming DM1 (DM2) and for both LT1 and LT2. 

According to requirement (i) above $\Rc=\Rc_\mathrm{NRR} > \Rc_\mathrm{H}$ must be the case; however, this is not satisfied at any radii. In other words, in order for the encounter to be sufficiently strong the orbit must be highly eccentric, but in this case the tidal force of the SBH would have stripped the star at an earlier epoch. For completeness we discuss the other requirement below. 

The debris disc can survive the encounters if $d_2<d_s$, where $d_s$ is the stripping distance, which can be estimated from requiring that the relative velocity $v_\mathrm{rel}$ between the star and planetesimal exceeds the escape velocity $v_\mathrm{esc}$ from the star. The former can be estimated as $v_\mathrm{rel} \sim (Gm_p/b^2)(2d/b)(2b/v_\mathrm{enc})$  \citep{ml12}, where $b$ is the impact parameter, $m_p$ is the perturber mass, assumed to be $m_p=m_\star=1\,\mathrm{M}_\odot$, and $v_\mathrm{enc}$ is the encounter velocity, which we approximate with the stellar velocity dispersion $\sigma$ (cf. equation~\ref{eq:jeans}); the latter is given by $v_\mathrm{esc} = (2Gm_\star/d)^{1/2}$. In terms of the change of the stellar speed the impact parameter can be estimated by $b \sim 2 Gm_p/(v_\mathrm{enc} \Delta v_a)$ where $\Delta v_a$ is the change of the apocentre speed (assuming that the encounters occur at apocentre). Setting $v_\mathrm{rel} = v_\mathrm{esc}$ subsequently gives the following relation for the stripping distance,
\begin{align}
d_s^3 \sim \frac{2(Gm_\star)^3}{v_\mathrm{enc}^2 \Delta v_a^4}.
\label{eq:d_s}
\end{align}
From the assumptions $\Delta r_a=0$ and $\Delta \Ec=0$, the definition of $\Rc$ and the relation $L=r_a v_a$, it follows that
\begin{align}
\Delta v_a(\Ec,\Rc) = \frac{L_\mathrm{c}(\Ec)}{2r_a(\Ec,\Rc)} \frac{\Delta \Rc}{\sqrt{\Rc}}.
\end{align}
Substituting $\Rc=\Rc_\mathrm{NRR}$ (cf. equation~\ref{eq:R_req_enc}) and using equation~(\ref{eq:r_p_change}) we find
\begin{align}
\Delta v_a(\Ec) = \frac{L_\mathrm{c}(\Ec)}{r_a(\Ec,\Rc_\mathrm{NRR})} \left ( \frac{P(\Ec)}{t_\Ec(\Ec)} \right )^{1/2}.
\label{eq:delta_v_a_enc}
\end{align}

Combining equations~(\ref{eq:d_s}) and (\ref{eq:delta_v_a_enc}) then gives the stripping distance $d_s$ as a function of $\Ec$, which is plotted in the bottom panel of Fig. \ref{fig:special_loss_cone_orbits} for LT1 and LT2. The outer disc radii in DM1 and DM2 are shown with horizontal lines. Destructive stripping, i.e. $d_s<d_2$, only occurs at a narrow radial range near $r\sim 1\,\mathrm{pc}$, therefore this does not pose a major problem in the scenario.

\section{Conclusions}
\label{sect:conclusions}
We have studied the dynamics of planetesimals in galactic nuclei focussing on the GC, and in the context of the tidal disruption of planetesimals by the SBH, possibly producing observable near infrared/X-ray flares. We assumed that the planetesimals were either formed in a large-scale cloud bound to the SBH, or in debris discs around stars. Our main conclusions are as follows.

(1) Assuming that planetesimals in the GC were initially formed in debris discs around stars, the tidal force of the SBH is effective at stripping the planetesimals from their parent stars at distances $r\lesssim 0.5 \,\ \mathrm{pc}$ from the SBH. Stripping by gravitational encounters with other stars is effective at stripping nearly all planetesimals within the radius of influence, $\approx 4 \, \mathrm{pc}$, after $\sim 100 \, \mathrm{Myr}$. 

(2) We studied the orbital evolution of planetesimals bound to the SBH in response to gravitational scattering from late-type stars, a hypothesized cusp of stellar black holes close to the SBH, and massive perturbers. We also included other effects such as physical collisions and resonant relaxation. We found that the disruption rate of planetesimals by the SBH at $t=10\, \mathrm{Gyr}$ is  $\approx 0.6\,\mathrm{day^{-1}}$, which is roughly consistent with the observed rate of the flares of once per day and the previous, less detailed, estimate by ZNM12, $\sim 1\,\mathrm{day^{-1}}$. Moreover, the rate is insensitive to model assumptions, in particular the initial distribution of planetesimals, i.e. whether the planetesimals were formed in a large-scale cloud or in debris discs around stars, and details of the gravitational perturbers. By comparing our rates in the case of formation in discs to the observed rates of \citealt{neilsen_ea13}, we have constrained the number of planetesimals per star to $N_\mathrm{a/\star} \approx 3.7 \times 10^7$; taking into account the effect of massive perturbers, we find $N_\mathrm{a/\star}\approx 1.1 \times 10^8$. 

(3) The result that both formation in a large-scale cloud and formation in debris discs around stars are consistent with the observed flaring rate suggests that it is not possible to distinguish between these two cases on the basis of the latter observation alone. However, the assumed number of planetesimals per star, $N_\mathrm{a/\star} = 2\times10^7$, is inferred from observations of debris discs around stars in the Solar neighbourhood. In the case of formation in a large-scale cloud this implies that the number of planetesimals formed is strongly correlated with the number of stars, and this requires finetuning of the quantity $N_\mathrm{a/\star}$. We favour the more natural explanation that planetesimals in galactic nuclei similar to the GC are formed no differently than planetesimals around stars in the Solar neighbourhood. 

(4) We have extrapolated our results to different galactic nuclei and we have estimated the event rate of the tidal disruption of planets by SBHs. Assuming one planet per star, we expect an observable planet disruption originating within the local Universe ($D<50\,\mathrm{Mpc}$) roughly every decade.

\section*{Acknowledgements}
We would like to thank S. Markoff and S. Nayakshin for enlightening discussions on the disruption of planetesimals in the Galactic Centre, and the referee, Hagai Perets, for providing very useful comments that helped to improve the paper. This work was supported by the Netherlands Research Council NWO (grants \#639.073.803 [VICI],  \#614.061.608 [AMUSE] and \#612.071.305 [LGM]) and the Netherlands Research School for Astronomy (NOVA).

\bibliographystyle{mn2e}
\bibliography{literature}

\begin{thebibliography}{76}
\expandafter\ifx\csname natexlab\endcsname\relax\def\natexlab#1{#1}\fi

\bibitem[{{Aharon} \& {Perets}(2014)}]{aharon_perets_14}
{Aharon} D., {Perets} H.~B., 2014, ArXiv e-prints

\bibitem[{{Alexander}(2005)}]{alexander05}
{Alexander} T., 2005, PhR, 419, 65

\bibitem[{{Alexander} \& {Pfuhl}(2014)}]{ap13}
{Alexander} T., {Pfuhl} O., 2014, ApJ, 780, 148

\bibitem[{{Antonini}(2014)}]{antonini14}
{Antonini} F., 2014, ArXiv e-prints

\bibitem[{{Antonini} \& {Merritt}(2013)}]{am13}
{Antonini} F., {Merritt} D., 2013, ApJL, 763, L10

\bibitem[{{Baganoff} {et~al}\mbox{.}(2001){Baganoff}, {Bautz}, {Brandt},
  {Chartas}, {Feigelson}, {Garmire}, {Maeda}, {Morris}, {Ricker}, {Townsley},
  \& {Walter}}]{baganoff_ea01}
{Baganoff} F.~K. {et~al.}, 2001, Natur, 413, 45

\bibitem[{{Baganoff} {et~al}\mbox{.}(2003){Baganoff}, {Maeda}, {Morris},
  {Bautz}, {Brandt}, {Cui}, {Doty}, {Feigelson}, {Garmire}, {Pravdo}, {Ricker},
  \& {Townsley}}]{baganoff_ea03}
{Baganoff} F.~K. {et~al.}, 2003, ApJ, 591, 891

\bibitem[{{Bahcall} \& {Wolf}(1976)}]{bw76}
{Bahcall} J.~N., {Wolf} R.~A., 1976, ApJ, 209, 214

\bibitem[{{Bahcall} \& {Wolf}(1977)}]{bw77}
{Bahcall} J.~N., {Wolf} R.~A., 1977, ApJ, 216, 883

\bibitem[{{Barri{\`e}re} {et~al}\mbox{.}(2014){Barri{\`e}re}, {Tomsick},
  {Baganoff}, {Boggs}, {Christensen}, {Craig}, {Dexter}, {Grefenstette},
  {Hailey}, {Harrison}, {Madsen}, {Mori}, {Stern}, {Zhang}, {Zhang}, \&
  {Zoglauer}}]{barriere_ea14}
{Barri{\`e}re} N.~M. {et~al.}, 2014, ApJ, 786, 46

\bibitem[{{Bartko} {et~al}\mbox{.}(2009){Bartko}, {Martins}, {Fritz}, {Genzel},
  {Levin}, {Perets}, {Paumard}, {Nayakshin}, {Gerhard}, {Alexander},
  {Dodds-Eden}, {Eisenhauer}, {Gillessen}, {Mascetti}, {Ott}, {Perrin},
  {Pfuhl}, {Reid}, {Rouan}, {Sternberg}, \& {Trippe}}]{bartko_ea09}
{Bartko} H. {et~al.}, 2009, ApJ, 697, 1741

\bibitem[{{Binney} \& {Tremaine}(2008)}]{bookbt08}
{Binney} J., {Tremaine} S., 2008, {Galactic Dynamics: Second Edition}.
  Princeton University Press

\bibitem[{{Blum} {et~al}\mbox{.}(2003){Blum}, {Ram{\'{\i}}rez}, {Sellgren}, \&
  {Olsen}}]{blum_ea03}
{Blum} R.~D., {Ram{\'{\i}}rez} S.~V., {Sellgren} K., {Olsen} K., 2003, ApJ,
  597, 323

\bibitem[{{Buchholz}, {Sch{\"o}del} \& {Eckart}(2009){Buchholz}, {Sch{\"o}del},
  \& {Eckart}}]{bse09}
{Buchholz} R.~M., {Sch{\"o}del} R., {Eckart} A., 2009, A\&A, 499, 483

\bibitem[{{Cohn}(1979)}]{cohn79}
{Cohn} H., 1979, ApJ, 234, 1036

\bibitem[{{Cohn} \& {Kulsrud}(1978)}]{ck78}
{Cohn} H., {Kulsrud} R.~M., 1978, ApJ, 226, 1087

\bibitem[{{Do} {et~al}\mbox{.}(2013){Do}, {Martinez}, {Yelda}, {Ghez},
  {Bullock}, {Kaplinghat}, {Lu}, {Peter}, \& {Phifer}}]{do_ea13}
{Do} T. {et~al.}, 2013, ApJL, 779, L6

\bibitem[{{Dodds-Eden} {et~al}\mbox{.}(2011){Dodds-Eden}, {Gillessen}, {Fritz},
  {Eisenhauer}, {Trippe}, {Genzel}, {Ott}, {Bartko}, {Pfuhl}, {Bower},
  {Goldwurm}, {Porquet}, {Trap}, \& {Yusef-Zadeh}}]{dodds-eden_ea11}
{Dodds-Eden} K. {et~al.}, 2011, ApJ, 728, 37

\bibitem[{{Eckart} {et~al}\mbox{.}(1993){Eckart}, {Genzel}, {Hofmann}, {Sams},
  \& {Tacconi-Garman}}]{eckart_ea93}
{Eckart} A., {Genzel} R., {Hofmann} R., {Sams} B.~J., {Tacconi-Garman} L.~E.,
  1993, ApJ, 407, L77

\bibitem[{{Eddington}(1916)}]{eddington16}
{Eddington} A.~S., 1916, MNRAS, 76, 572

\bibitem[{{Eilon}, {Kupi} \& {Alexander}(2009){Eilon}, {Kupi}, \&
  {Alexander}}]{eka09}
{Eilon} E., {Kupi} G., {Alexander} T., 2009, ApJ, 698, 641

\bibitem[{{Eisenhauer} {et~al}\mbox{.}(2005){Eisenhauer}, {Genzel},
  {Alexander}, {Abuter}, {Paumard}, {Ott}, {Gilbert}, {Gillessen}, {Horrobin},
  {Trippe}, {Bonnet}, {Dumas}, {Hubin}, {Kaufer}, {Kissler-Patig}, {Monnet},
  {Str{\"o}bele}, {Szeifert}, {Eckart}, {Sch{\"o}del}, \&
  {Zucker}}]{eisenhauer_ea05}
{Eisenhauer} F. {et~al.}, 2005, ApJ, 628, 246

\bibitem[{{Ferrarese} \& {Merritt}(2000)}]{ferrarese_merritt00}
{Ferrarese} L., {Merritt} D., 2000, ApJL, 539, L9

\bibitem[{{Frank} \& {Rees}(1976)}]{fr76}
{Frank} J., {Rees} M.~J., 1976, MNRAS, 176, 633

\bibitem[{{Genzel}, {Eisenhauer} \& {Gillessen}(2010){Genzel}, {Eisenhauer}, \&
  {Gillessen}}]{geg10}
{Genzel} R., {Eisenhauer} F., {Gillessen} S., 2010, Reviews of Modern Physics,
  82, 3121

\bibitem[{{Genzel} {et~al}\mbox{.}(2003{\natexlab{a}}){Genzel}, {Sch{\"o}del},
  {Ott}, {Eckart}, {Alexander}, {Lacombe}, {Rouan}, \&
  {Aschenbach}}]{genzel_ea03}
{Genzel} R., {Sch{\"o}del} R., {Ott} T., {Eckart} A., {Alexander} T., {Lacombe}
  F., {Rouan} D., {Aschenbach} B., 2003{\natexlab{a}}, Natur, 425, 934

\bibitem[{{Genzel} {et~al}\mbox{.}(2003{\natexlab{b}}){Genzel}, {Sch{\"o}del},
  {Ott}, {Eisenhauer}, {Hofmann}, {Lehnert}, {Eckart}, {Alexander},
  {Sternberg}, {Lenzen}, {Cl{\'e}net}, {Lacombe}, {Rouan}, {Renzini}, \&
  {Tacconi-Garman}}]{genzel_ea03b}
{Genzel} R. {et~al.}, 2003{\natexlab{b}}, ApJ, 594, 812

\bibitem[{{Genzel} {et~al}\mbox{.}(1996){Genzel}, {Thatte}, {Krabbe}, {Kroker},
  \& {Tacconi-Garman}}]{genzel_ea96}
{Genzel} R., {Thatte} N., {Krabbe} A., {Kroker} H., {Tacconi-Garman} L.~E.,
  1996, ApJ, 472, 153

\bibitem[{{Ghez} {et~al}\mbox{.}(2008){Ghez}, {Salim}, {Weinberg}, {Lu}, {Do},
  {Dunn}, {Matthews}, {Morris}, {Yelda}, {Becklin}, {Kremenek},
  {Milosavljevic}, \& {Naiman}}]{ghez_ea08}
{Ghez} A.~M. {et~al.}, 2008, ApJ, 689, 1044

\bibitem[{{Gillessen} {et~al}\mbox{.}(2009){Gillessen}, {Eisenhauer}, {Trippe},
  {Alexander}, {Genzel}, {Martins}, \& {Ott}}]{gillessen_ea09}
{Gillessen} S., {Eisenhauer} F., {Trippe} S., {Alexander} T., {Genzel} R.,
  {Martins} F., {Ott} T., 2009, ApJ, 692, 1075

\bibitem[{{Gomes} {et~al}\mbox{.}(2005){Gomes}, {Levison}, {Tsiganis}, \&
  {Morbidelli}}]{gomes_ea_05}
{Gomes} R., {Levison} H.~F., {Tsiganis} K., {Morbidelli} A., 2005, Nature, 435,
  466

\bibitem[{{G{\"u}rkan} \& {Hopman}(2007)}]{gh07}
{G{\"u}rkan} M.~A., {Hopman} C., 2007, MNRAS, 379, 1083

\bibitem[{Guyer, Wheeler \& Warren(2009)Guyer, Wheeler, \& Warren}]{gww09}
Guyer J.~E., Wheeler D., Warren J.~A., 2009, Computing in Science \&
  Engineering, 11, 6

\bibitem[{{Hamers}, {Portegies Zwart} \& {Merritt}(2014){Hamers}, {Portegies
  Zwart}, \& {Merritt}}]{hpm14}
{Hamers} A.~S., {Portegies Zwart} S.~F., {Merritt} D., 2014, ArXiv e-prints

\bibitem[{{Hamilton} \& {Burns}(1992)}]{hb92}
{Hamilton} D.~P., {Burns} J.~A., 1992, Icar, 96, 43

\bibitem[{{Hayashi}(1981)}]{hayashi81}
{Hayashi} C., 1981, Progress of Theoretical Physics Supplement, 70, 35

\bibitem[{{Heggie}(1975)}]{heggie75}
{Heggie} D.~C., 1975, MNRAS, 173, 729

\bibitem[{{Heggie}, {Hut} \& {McMillan}(1996){Heggie}, {Hut}, \&
  {McMillan}}]{heggiehut96}
{Heggie} D.~C., {Hut} P., {McMillan} S.~L.~W., 1996, ApJ, 467, 359

\bibitem[{{Hills}(1988)}]{hills88}
{Hills} J.~G., 1988, Natur, 331, 687

\bibitem[{{Hopman} \& {Alexander}(2006)}]{ha06}
{Hopman} C., {Alexander} T., 2006, ApJ, 645, 1152

\bibitem[{{Hurley}, {Pols} \& {Tout}(2000){Hurley}, {Pols}, \& {Tout}}]{hpt00}
{Hurley} J.~R., {Pols} O.~R., {Tout} C.~A., 2000, MNRAS, 315, 543

\bibitem[{{Hut}(1983)}]{hut83}
{Hut} P., 1983, ApJ, 268, 342

\bibitem[{{Hut}(1993)}]{hut93}
{Hut} P., 1993, ApJ, 403, 256

\bibitem[{{Hut} \& {Bahcall}(1983)}]{hutbahcall83}
{Hut} P., {Bahcall} J.~N., 1983, ApJ, 268, 319

\bibitem[{{Hut}, {Makino} \& {McMillan}(1995){Hut}, {Makino}, \&
  {McMillan}}]{hmm95}
{Hut} P., {Makino} J., {McMillan} S., 1995, ApJL, 443, L93

\bibitem[{{Lestrade} {et~al}\mbox{.}(2011){Lestrade}, {Morey}, {Lassus}, \&
  {Phou}}]{lmlp11}
{Lestrade} J.-F., {Morey} E., {Lassus} A., {Phou} N., 2011, A\&A, 532, A120

\bibitem[{{Lu} {et~al}\mbox{.}(2009){Lu}, {Ghez}, {Hornstein}, {Morris},
  {Becklin}, \& {Matthews}}]{lu_ea09}
{Lu} J.~R., {Ghez} A.~M., {Hornstein} S.~D., {Morris} M.~R., {Becklin} E.~E.,
  {Matthews} K., 2009, ApJ, 690, 1463

\bibitem[{{Maness} {et~al}\mbox{.}(2007){Maness}, {Martins}, {Trippe},
  {Genzel}, {Graham}, {Sheehy}, {Salaris}, {Gillessen}, {Alexander}, {Paumard},
  {Ott}, {Abuter}, \& {Eisenhauer}}]{maness_ea07}
{Maness} H. {et~al.}, 2007, ApJ, 669, 1024

\bibitem[{{Merritt}(2004)}]{merritt04}
{Merritt} D., 2004, Physical Review Letters, 92, 201304

\bibitem[{{Merritt}(2010)}]{merritt10}
{Merritt} D., 2010, ApJ, 718, 739

\bibitem[{{Merritt}(2013)}]{bookmerritt13}
{Merritt} D., 2013, {Dynamics and Evolution of Galactic Nuclei}

\bibitem[{{Merritt} {et~al}\mbox{.}(2011){Merritt}, {Alexander}, {Mikkola}, \&
  {Will}}]{mamw11}
{Merritt} D., {Alexander} T., {Mikkola} S., {Will} C.~M., 2011, PhRvD, 84,
  044024

\bibitem[{{Merritt}, {Harfst} \& {Bertone}(2007){Merritt}, {Harfst}, \&
  {Bertone}}]{mhb07}
{Merritt} D., {Harfst} S., {Bertone} G., 2007, PhRvD, 75, 043517

\bibitem[{{Murray-Clay} \& {Loeb}(2012)}]{ml12}
{Murray-Clay} R.~A., {Loeb} A., 2012, Natur, 3

\bibitem[{{Nayakshin}, {Sazonov} \& {Sunyaev}(2012){Nayakshin}, {Sazonov}, \&
  {Sunyaev}}]{nss12}
{Nayakshin} S., {Sazonov} S., {Sunyaev} R., 2012, MNRAS, 419, 1238

\bibitem[{{Neilsen} {et~al}\mbox{.}(2013){Neilsen}, {Nowak}, {Gammie},
  {Dexter}, {Markoff}, {Haggard}, {Nayakshin}, {Wang}, {Grosso}, {Porquet},
  {Tomsick}, {Degenaar}, {Fragile}, {Houck}, {Wijnands}, {Miller}, \&
  {Baganoff}}]{neilsen_ea13}
{Neilsen} J. {et~al.}, 2013, ApJ, 774, 42

\bibitem[{{Niko{\l}ajuk} \& {Walter}(2013)}]{nw13}
{Niko{\l}ajuk} M., {Walter} R., 2013, A\&A, 552, A75

\bibitem[{{Oh}, {Kim} \& {Figer}(2009){Oh}, {Kim}, \& {Figer}}]{ohf09}
{Oh} S., {Kim} S.~S., {Figer} D.~F., 2009, Journal of Korean Astronomical
  Society, 42, 17

\bibitem[{{Paumard} {et~al}\mbox{.}(2006){Paumard}, {Genzel}, {Martins},
  {Nayakshin}, {Beloborodov}, {Levin}, {Trippe}, {Eisenhauer}, {Ott},
  {Gillessen}, {Abuter}, {Cuadra}, {Alexander}, \& {Sternberg}}]{paumard_ea06}
{Paumard} T. {et~al.}, 2006, Journal of Physics Conference Series, 54, 199

\bibitem[{{Pelupessy} {et~al}\mbox{.}(2013){Pelupessy}, {van Elteren}, {de
  Vries}, {McMillan}, {Drost}, \& {Portegies Zwart}}]{pelupessy_ea13}
{Pelupessy} F.~I., {van Elteren} A., {de Vries} N., {McMillan} S.~L.~W.,
  {Drost} N., {Portegies Zwart} S.~F., 2013, A\&A, 557, A84

\bibitem[{{Perets}(2009)}]{perets_09}
{Perets} H.~B., 2009, ApJ, 690, 795

\bibitem[{{Perets} {et~al}\mbox{.}(2009){Perets}, {Gualandris}, {Kupi},
  {Merritt}, \& {Alexander}}]{perets_ea_09}
{Perets} H.~B., {Gualandris} A., {Kupi} G., {Merritt} D., {Alexander} T., 2009,
  ApJ, 702, 884

\bibitem[{{Perets}, {Hopman} \& {Alexander}(2007){Perets}, {Hopman}, \&
  {Alexander}}]{pha07}
{Perets} H.~B., {Hopman} C., {Alexander} T., 2007, ApJ, 656, 709

\bibitem[{{Pfuhl} {et~al}\mbox{.}(2011){Pfuhl}, {Fritz}, {Zilka}, {Maness},
  {Eisenhauer}, {Genzel}, {Gillessen}, {Ott}, {Dodds-Eden}, \&
  {Sternberg}}]{pfuhl_ea11}
{Pfuhl} O. {et~al.}, 2011, ApJ, 741, 108

\bibitem[{{Porquet} {et~al}\mbox{.}(2003){Porquet}, {Predehl}, {Aschenbach},
  {Grosso}, {Goldwurm}, {Goldoni}, {Warwick}, \& {Decourchelle}}]{porquet_ea03}
{Porquet} D., {Predehl} P., {Aschenbach} B., {Grosso} N., {Goldwurm} A.,
  {Goldoni} P., {Warwick} R.~S., {Decourchelle} A., 2003, A\&A, 407, L17

\bibitem[{{Portegies Zwart} {et~al}\mbox{.}(2013){Portegies Zwart}, {McMillan},
  {van Elteren}, {Pelupessy}, \& {de Vries}}]{pmepv13}
{Portegies Zwart} S., {McMillan} S.~L.~W., {van Elteren} E., {Pelupessy} I.,
  {de Vries} N., 2013, Computer Physics Communications, 183, 456

\bibitem[{{Rauch} \& {Tremaine}(1996)}]{rt96}
{Rauch} K.~P., {Tremaine} S., 1996, NewA, 1, 149

\bibitem[{Rosenbluth, MacDonald \& Judd(1957)Rosenbluth, MacDonald, \&
  Judd}]{rmj57}
Rosenbluth M.~N., MacDonald W.~M., Judd D.~L., 1957, Phys. Rev., 107, 1

\bibitem[{{Sch{\"o}del} {et~al}\mbox{.}(2007){Sch{\"o}del}, {Eckart},
  {Alexander}, {Merritt}, {Genzel}, {Sternberg}, {Meyer}, {Kul}, {Moultaka},
  {Ott}, \& {Straubmeier}}]{schodel_ea07}
{Sch{\"o}del} R. {et~al.}, 2007, A\&A, 469, 125

\bibitem[{{Sch{\"o}del} {et~al}\mbox{.}(2014){Sch{\"o}del}, {Feldmeier},
  {Kunneriath}, {Stolovy}, {Neumayer}, {Amaro-Seoane}, \&
  {Nishiyama}}]{schodel_ea14}
{Sch{\"o}del} R., {Feldmeier} A., {Kunneriath} D., {Stolovy} S., {Neumayer} N.,
  {Amaro-Seoane} P., {Nishiyama} S., 2014, A\&A, 566, A47

\bibitem[{{Sch{\"o}del}, {Merritt} \& {Eckart}(2009){Sch{\"o}del}, {Merritt},
  \& {Eckart}}]{schodel_ea09}
{Sch{\"o}del} R., {Merritt} D., {Eckart} A., 2009, A\&A, 502, 91

\bibitem[{{Shen} {et~al}\mbox{.}(2005){Shen}, {Lo}, {Liang}, {Ho}, \&
  {Zhao}}]{shen_ea05}
{Shen} Z.-Q., {Lo} K.~Y., {Liang} M.-C., {Ho} P.~T.~P., {Zhao} J.-H., 2005,
  Natur, 438, 62

\bibitem[{{Trippe} {et~al}\mbox{.}(2008){Trippe}, {Gillessen}, {Gerhard},
  {Bartko}, {Fritz}, {Maness}, {Eisenhauer}, {Martins}, {Ott}, {Dodds-Eden}, \&
  {Genzel}}]{trippe_ea08}
{Trippe} S. {et~al.}, 2008, A\&A, 492, 419

\bibitem[{{Vika} {et~al}\mbox{.}(2009){Vika}, {Driver}, {Graham}, \&
  {Liske}}]{vika_ea09}
{Vika} M., {Driver} S.~P., {Graham} A.~W., {Liske} J., 2009, MNRAS, 400, 1451

\bibitem[{{Wyatt}(2008)}]{wyatt08}
{Wyatt} M.~C., 2008, ARA\&A, 46, 339

\bibitem[{{Zubovas}, {Nayakshin} \& {Markoff}(2012){Zubovas}, {Nayakshin}, \&
  {Markoff}}]{znm12}
{Zubovas} K., {Nayakshin} S., {Markoff} S., 2012, MNRAS, 421, 1315

\end{thebibliography}

\appendix
\onecolumn

\section{Planetesimal stripping}
\label{app:fstrip}
\subsection{Stripping by the SBH}
\label{app:fstrip:SBH}
In the middle panel of Fig. (\ref{fig:stripping_fraction_SBH}) the stripping fraction $f_\mathrm{strip}$ is expressed in terms of the pericentre distance $r_p$ of the orbit of the star around the SBH. Here, we discuss our method to express $f_\mathrm{strip}$ in terms of the orbital energy, assuming an isotropic velocity distribution. 

First we express $r_p$ in terms of energy $\Ec$ and $\Rc$, where angular momentum $L$ is expressed in terms of $\Rc\equiv (L/L_\mathrm{c})^2\in[0,1]$; here, $L_\mathrm{c}(\Ec)$ is the angular momentum of a circular orbit with energy $\Ec$. The latter is given in terms of the potential by $L_\mathrm{c}^2=-r_c^3 \, \mathrm{d} \psi/\mathrm{d} r|_{r=r_c}$, with $r_c=r_c(\mathcal{E})$ the radius of a circular orbit with energy $\mathcal{E}$, which is the solution of the equation $2[\psi(r_c)-\mathcal{E}] + r_c \, \mathrm{d} \psi/\mathrm{d} r|_{r=r_c} = 0$ (e.g. \citealt{cohn79}; \citealt[][5.5.1]{bookmerritt13}). We determine the pericentre distance $r_p$ by finding the smallest solution $r=r(\mathcal{E},\Rc)$ for which the radial velocity $v_r=v_r(r,\mathcal{E},\Rc) = \{ 2[\psi(r)-\mathcal{E}] - \Rc \,L_\mathrm{c}^2(\mathcal{E})/r^2 \}^{1/2}$ vanishes. From $r_p(\Ec,\Rc)$ we compute the stripping fraction $f_\mathrm{strip}(\Ec,\Rc)$. Subsequently, we average this quantity over $\Rc$ assuming an isotropic velocity distribution, $f_\mathrm{strip}(\mathcal{E}) \equiv \int_{\Rc_\mathrm{lc}(\mathcal{E})}^1 \mathrm{d} \Rc\, f_\mathrm{strip}(\mathcal{E},\Rc)$. Here, $\Rc_\mathrm{lc}(\mathcal{E})$ is the value of $\Rc$ that corresponds to the loss cone $r_\mathrm{lc}$ of the star; we adopt $r_\mathrm{lc} = 1\,\mathrm{AU}$, approximately the tidal disruption radius of a solar-type star. An orbit with energy $\Ec$ and $\Rc=\Rc_\mathrm{lc}$ just grazing the loss cone $r_\mathrm{lc}$ at pericentre has $v_r(r_\mathrm{lc},\Ec,\Rc_\mathrm{lc})=v_r(r_p,\Ec,R_\mathrm{lc})=0$, which implies $\Rc_\mathrm{lc}=[2r_\mathrm{lc}^2/L_\mathrm{c}^2(\Ec)][\psi(r_\mathrm{lc})-\Ec]$. 

The resulting stripping fractions $f_\mathrm{strip}(\Ec)$ are plotted in the bottom panel of Fig. \ref{fig:stripping_fraction_SBH} with solid lines. Analytic expressions for $f_\mathrm{strip}(\Ec)$ can be derived if the stellar potential is neglected. In that case the pericentre distance $r_p(\Ec,\Rc) = [GM_\bullet/(2\Ec)][1-\sqrt{1-\Rc}] = a(1-e)$ and $d_\mathrm{strip} = a f_\mathrm{H} (1-e)$, where $f_\mathrm{H} \equiv [m_\star/(3M_\bullet)]^{1/3}$ (cf. equation~\ref{eq:rstripSBH}). Using equation~\ref{eq:fstrip} and assuming that $\Rc_\mathrm{lc} = 0$, we find for the angular-momentum-averaged stripping fraction $f_\mathrm{strip}(a) \approx \int_0^1 \,\mathrm{d} \Rc \, f_\mathrm{strip}(\Ec,\Rc) = \int_0^1 \mathrm{d}e \, 2e f_\mathrm{strip}(a,e)$, for $a f_\mathrm{H} > d_2$,
\begin{align}
\nonumber f_\mathrm{strip}(a) &= \frac{1}{d_2^{2-\beta}-d_1^{2-\beta}} \left \{ d_2^{2-\beta} \left [ \left (1-\frac{d_1}{a f_\mathrm{H}} \right )^2 -  \left (1-\frac{d_2}{a f_\mathrm{H}} \right )^2 \right ]  - 2 \left [ (a f_\mathrm{H})^{2+\beta} (\beta-4)(\beta-3) \right ]^{-1} \right.\\
\nonumber &\quad \times \left. \left [ d_1^3 \left ( \frac{d_1}{a f_\mathrm{H}} \right )^{-\beta} \left( a f_\mathrm{H}(\beta-4) - d_1(\beta-3) \right ) - d_2^3 \left ( \frac{d_2}{a f_\mathrm{H}} \right )^{-\beta} \left( a f_\mathrm{H}(\beta-4) - d_2(\beta-3) \right ) \right ] \right \} + 1 - \left(1-\frac{d_1}{a f_\mathrm{H}} \right )^2;
\end{align}
for $d_1 \leq a f_\mathrm{H} \leq d_2$,
\begin{align}
\nonumber f_\mathrm{strip}(a) &= \frac{1}{d_2^{2-\beta}-d_1^{2-\beta}} \left \{ d_2^{2-\beta} \left (1-\frac{d_1}{a f_\mathrm{H}} \right )^2   - 2 (a f_\mathrm{H})^{2-\beta} \left [ 
\frac{ d_1^3 \left ( \frac{d_1}{a f_\mathrm{H}} \right )^{-\beta} \left( a f_\mathrm{H}(\beta-4) - d_1(\beta-3) \right ) }{(a f_\mathrm{H})^4 (\beta-4)(\beta-3)} + \frac{1}{12 - 7\beta+\beta^2} \right ] \right \} \\
& \quad + 1 - \left(1-\frac{d_1}{a f_\mathrm{H}} \right )^2
\end{align}
and $f_\mathrm{strip}(a) = 1$ for $a f_\mathrm{H} < d_1$. The stripping fractions according to this approximation are shown with dashed lines in the bottom panel of Fig. \ref{fig:stripping_fraction_SBH}.

\subsection{Stripping by gravitational encounters}
\label{app:fstrip:enc}
We compute the orbital average of the local stripping fraction arising from gravitational encounters using the relation (e.g. \citealt[][5.5.2]{bookmerritt13})
\begin{align}
f_\mathrm{strip}(\Ec) = \frac{4}{p(\Ec)} \int_0^{\psi^{-1}(\Ec)} \mathrm{d} r \, r^2 v(r,\Ec) \, f_\mathrm{strip}(r),
\label{eq:f_strip_enc_orb_av}
\end{align}
where $p(\Ec)$ is the phase space volume per unit energy and is given by
\begin{align}
p(\Ec) = 4 \int_0^{\psi^{-1}(\Ec)} \, \mathrm{d} r \, r^2 v(r,\Ec),
\label{eq:p_e}
\end{align}
with $\psi^{-1}(\Ec)$ the inverse function of $\psi(r)$ and $v(r,\Ec) = \sqrt{2[\psi(r)-\Ec]}$ the orbital speed at radius $r$ for an orbit with energy $\Ec$. In equation~(\ref{eq:f_strip_enc_orb_av}) an isotropic velocity distribution is assumed; it was derived from the more general expression for the orbital average of a local function $f(r)$ (e.g. \citealt{ck78}), 
\begin{align}
f (\Ec,\Rc) = \frac{2}{P(\Ec,\Rc)} \int_{r_{-}(\Ec,\Rc)}^{r_{+}(\Ec,\Rc)} \frac{\mathrm{d} r}{v_r(r,\Ec,\Rc)} f(r),
\label{eq:f_strip_enc_orb_av_R}
\end{align}
where $P(\Ec,\Rc) = 2 \int_{r_{-}}^{r_{+}} \mathrm{d} r/v_r$ is the orbital period and $r_\pm(\Ec,\Rc)$ are the turning points of the orbit, i.e. the solutions to $v_r(r_\pm,\Ec,R)=0$. In this derivation several approximations have been made. To verify the validity of the simplified orbit average, equation~(\ref{eq:f_strip_enc_orb_av}), we have computed $f_\mathrm{strip}(\Ec)$ numerically using both equations~(\ref{eq:f_strip_enc_orb_av}) and (\ref{eq:f_strip_enc_orb_av_R}), in the latter case averaging over $R$ after calculating $f(\Ec,R)$. We find that there is no discernable difference in the resulting averaged function $f_\mathrm{strip}(\Ec)$ between the two cases.

\section{Terms appearing in the Fokker-Planck equation}
\label{app:FP_main}

\subsection{Gravitational scattering flux}
\label{app:FP_main:scat}
The quantity $\mathcal{F}_\Ec$ represents a flux in $\Ec$-space because of gravitational scattering by massive scatterers and is given by (e.g. \citealt[][7.1.2.2]{bookmerritt13})
\begin{align}
\mathcal{F}_\Ec(\Ec,t) = D_{\Ec\Ec}(\Ec) \frac{\partial f_\mathrm{a}(\Ec,t)}{\partial \Ec},
\end{align}
where $D_{\Ec\Ec}(\Ec)$ is an energy diffusion coefficient that depends on the distribution function of the scatterers. Assuming a discrete mass spectrum of scatterers, $D_{\Ec\Ec}$ is given by
\begin{align}
\nonumber D_{\Ec\Ec}(\Ec) &= \sum_j m_j^2 \ln(\Lambda_j) D_{\Ec\Ec;j} = 64 \pi^5 G^2 \sum_j m_j^2 \ln(\Lambda_j) \\
& \times \left [ q(\Ec) \int_{-\infty}^\Ec \mathrm{d} \Ec' \, f_j (\Ec') + \int_\Ec^{\infty} \mathrm{d} \Ec' \, f_j (\Ec') q (\Ec') \right ],
\label{eq:D_ee}
\end{align}
where $\ln(\Lambda_j)$ is the Coulomb logarithm, $f_j(\Ec)$ is the distribution function of massive scatterer $j$ with mass $m_j$ and
\begin{align}
q(\Ec) = \frac{4}{3} \int_0^{\psi^{-1}(\Ec)} \, \mathrm{d} r \, r^2 v^3(r,\Ec).
\label{eq:q_e}
\end{align}
Here, we adopt $\ln(\Lambda_j)=\ln[M_\bullet/(2m_j)]$. The distribution functions $f_j(\Ec)$ are computed from the number density $n_j(r)$ and potential $\psi(r)$ using Eddington's formula \citep{eddington16}, 
\begin{align}
f_j(\Ec) = \frac{\sqrt{2}}{4\pi^2} \frac{\partial}{\partial \Ec} \int_{-\infty}^{\psi^{-1}(\Ec)} \frac{\mathrm{d}r}{\sqrt{\Ec-\psi(r)}} \frac{\mathrm{d}n_j(r)}{\mathrm{d}r}.
\label{eq:distr_edd}
\end{align}

\subsubsection{Black hole cusp}
\label{app:FP_main:scat:bh_cusp}
In the case of a cusp of stellar black holes (cf. Section~\ref{sect:models:GC:stellar_black_holes}) we derive and adopt semi-analytic expressions for $D_{\Ec \Ec;\mathrm{BH}}(\Ec)$. We assume that the distribution function $f_\mathrm{BH}(\Ec)$ has the form of a truncated power-law,
\begin{align}
f_\mathrm{BH}(\Ec) = \left \{ \begin{array}{cc}
f_0 \Ec^{\gamma-3/2}, & \Ec > \Ec_d; \\
0, & \Ec \leq \Ec_d.
\end{array} \right.
\end{align}
Here, $\Ec_d \equiv GM_\bullet/(2r_d)$ and $r_d=0.2\,\mathrm{pc}$ is assumed; furthermore we adopt $\gamma=1.8$. The quantity $f_0$ can be expressed in terms of the number density of stellar black holes at $r=r_d$, $n_d = N_\mathrm{BH}(3-\gamma)/(4\pi r_d^3)$, where $N_\mathrm{BH}=4800$ is the total number of black holes within $r_d$, via
\begin{align}
f_0 = \frac{\sqrt{2}}{4} \pi^{-3/2} \frac{\Gamma(\gamma+1)}{\Gamma(\gamma-\frac{1}{2})} \left ( \frac{r_d}{GM_\bullet} \right )^\gamma n_d,
\label{eq:f0_n0}
\end{align}
where $\Gamma(x)$ is the Gamma function. For $r\leq r_d$ it is well-justified to neglect the stellar potential in $\psi(r)$. With this approximation equation~(\ref{eq:D_ee}) yields
\begin{align}
D_{\Ec\Ec;\mathrm{BH}}(\Ec) = \frac{32}{3} \sqrt{2} \pi^6 G^2 (GM_\bullet)^3 f_0 \frac{1}{(2-\gamma)(2\gamma-1)} \times \left \{ \begin{array}{cc}
3\Ec^{\gamma-2} - 2(2-\gamma)\Ec_d^{\gamma-1/2} \Ec^{-3/2}, & \Ec > \Ec_d; \\
(2\gamma-1)\Ec_d^{\gamma-2}, & \Ec \leq \Ec_d.
\end{array} \right.
\end{align}

\subsection{Flux into the loss cone}
\label{app:FP_main:lc}
As discussed in Section~\ref{sect:dyn_ff_ast:eq}, relaxation through gravitational encounters typically occurs much faster in angular-momentum space than in energy space. Scattering into the loss cone of the SBH is therefore dominated by diffusion in angular momentum rather than diffusion in energy. The former is described in equation~(\ref{eq:FP_main}) by the term $F_\mathrm{lc}(\Ec,t)$, the flux in angular-momentum space into the loss cone. We adopt the formalism of the Cohn-Kulsrud boundary layer, which is based on the Fokker-Planck equation in angular-momentum space (\citealt{ck78},\citealt[][6.1.2]{bookmerritt13}). By solving this equation both in local and orbit-averaged form and matching the two solutions, the following expression can be derived for the flux into the loss cone,
\begin{align}
F_\mathrm{lc}(\Ec,t) \approx 4\pi^2 P_r(\Ec) L_\mathrm{c}^2(\Ec) \bar{\mu}(\Ec) \left \{ \ln \left [\Rc_0(\Ec)^{-1} \right ] \right \}^{-1} f_\mathrm{a}(\Ec,t).
\label{eq:F_lc}
\end{align}
Here, $P_r(\Ec)=\int_0^{\psi^{-1}(\Ec)} \mathrm{d} r/v_r(r,\Ec,0)$ is the radial orbital period and $\bar{\mu}(\Ec)$ is an orbit-averaged angular-momentum diffusion coefficient in the limit $\Rc\rightarrow0$, defined by
\begin{align}
\bar{\mu}(\Ec) &= \frac{2}{P_r(\Ec)} \int_0^{\psi^{-1}(\Ec)} \frac{\mathrm{d}r}{v_r(r,\Ec,0)} \lim_{\Rc\rightarrow0} \frac{ \langle (\Delta \Rc)^2 \rangle}{(2\Rc)},
\end{align}
where $\langle (\Delta \Rc)^2 \rangle$ is the second-order diffusion coefficient in $\Rc$ arising from NRR. Using standard expressions for $\langle (\Delta \Rc)^2 \rangle$ (e.g. \citealt{ck78}), $\bar{\mu}(\Ec)$ can be expressed in terms of $f_j$ as
\begin{align}
\bar{\mu}(\Ec) &= \frac{32 \pi^2 G^2}{3P_r(\Ec)L_\mathrm{c}^2(\Ec)} \sum_j m_j^2 \ln(\Lambda_j) \left [ 3 \bar{I}_{1/2;j}(\Ec) + 2 \bar{I}_{0;j}(\Ec) - \bar{I}_{3/2;j}(\Ec) \right ],
\label{eq:mu_e}
\end{align}
where 
\begin{align}
\nonumber \bar{I}_{0;j}(\Ec) &= \int_0^{\psi^{-1}(\Ec)} \frac{\mathrm{d} r \, r^2}{\sqrt{2[\psi(r)-\Ec)]}} \int_{-\infty}^\Ec \mathrm{d} \Ec' \, f_j (\Ec'); \\
\bar{I}_{n/2;j}(\Ec) &= \int_0^{\psi^{-1}(\Ec)} \frac{\mathrm{d} r \, r^2}{\sqrt{2[\psi(r)-\Ec]}} \int_{\Ec}^{\psi(r)} \mathrm{d} \Ec' \, \left ( \frac{\psi(r)-\Ec'}{\psi(r)-\Ec} \right )^{n/2} f_j (\Ec' ).
\label{eq:mu_e_I}
\end{align}

In equation~(\ref{eq:F_lc}) it is assumed that the distribution function $f_\mathrm{a}$ decreases logarithmically with $\Rc$ as $\Rc\rightarrow0$. The value of $\Rc$ for which $f_\mathrm{a}$ vanishes is not the capture boundary $\Rc_\mathrm{lc}(\Ec)$: far away from the SBH, where the orbital period is relatively long compared to the angular-momentum relaxation time-scale (the `full loss cone' regime), a planetesimal can be scattered into and out of the loss cone without being immediately disrupted, i.e. it may be that $f_\mathrm{a}(\Ec,\Rc_\mathrm{lc})>0$. The value of $\Rc$  for which the distribution function does vanish is given by the function $\Rc_0(\Ec)= \Rc_\mathrm{lc}(\Ec) \exp[-\tilde{q}/\xi(\tilde{q})]<\Rc_\mathrm{lc}$, where $\tilde{q}=\tilde{q}(\Ec) = P_r(\Ec) \bar{\mu}(\Ec)/\Rc_\mathrm{lc}(\Ec)$ and $\xi(\tilde{q})$ is given by
\begin{align}
\nonumber \xi(\tilde{q}) \equiv 1 - \sum_{m=1}^\infty \frac{\exp \left (-\alpha_m^2 \tilde{q}/4 \right )}{\alpha_m^2},
\end{align}
where $\alpha_m$ is the $m^\mathrm{th}$ zero of the Bessel function $J_0(\alpha)$ of the first kind. We compute the quantity $\Rc_\mathrm{lc}$ from $\Rc_\mathrm{lc} = 2[r_\mathrm{lc}^2/L_\mathrm{c}^2(\Ec)][\psi(r_\mathrm{lc}) - \Ec]$ with $r_\mathrm{lc} = 1 \, \mathrm{AU}$. 

\subsubsection{Black hole cusp}
\label{app:FP_main:lc:bh_cusp}
Similarly as in Section~\ref{app:FP_main:scat:bh_cusp} we derive and adopt semi-analytic expressions for $\overline{\mu}(\Ec)$ for the case of a cusp of stellar black holes. With the same assumptions as in Section~\ref{app:FP_main:scat:bh_cusp} the loss cone integrals $\bar{I}_{0;\mathrm{BH}}$ and $\bar{I}_{n/2;\mathrm{BH}}$ (cf. equation~\ref{eq:mu_e_I}) can be expressed as
\begin{align}
\nonumber \bar{I}_{0;\mathrm{BH}}(\Ec) &= \frac{5\sqrt{2} \pi}{16} \frac{f_0}{2\gamma-1} (GM_\bullet)^3 \times \left \{ \begin{array}{cc}
\Ec^{\gamma-4} - \Ec_d^{\gamma-1/2} \Ec^{-7/2}, & \Ec>\Ec_d ; \\
0, & \Ec \leq \Ec_d; 
\end{array} \right. \\
\bar{I}_{n/2;\mathrm{BH}}(\Ec) &= \frac{\sqrt{2}}{2n+1} f_0 (GM_\bullet)^3 \times \left \{ \begin{array}{cc}
g_n(\Ec,\Ec,\gamma), & \Ec > \Ec_d; \\
g_n(\Ec,\Ec_d,\gamma), & \Ec \leq \Ec_d,
\end{array} \right. \\
\end{align}
where 
\begin{align}
\nonumber g_1(\Ec,\Ec',\gamma) &= \Ec'^{\gamma-4} \int_0^1 \mathrm{d} x \, x^{3-\gamma} \left (1- \frac{\Ec}{\Ec'} x \right)^{-1} (1-x)^{3/2} \, {}_2F_1 \left ( \frac{3}{2},\frac{3}{2}-\gamma,\frac{5}{2};1-x\right); \\
\nonumber g_3(\Ec,\Ec',\gamma) &= \Ec'^{\gamma-4} \int_0^1 \mathrm{d} x \, x^{3-\gamma} \left (1- \frac{\Ec}{\Ec'} x \right)^{-2} (1-x)^{5/2} \, {}_2F_1 \left ( \frac{5}{2},\frac{3}{2}-\gamma,\frac{7}{2};1-x\right).
\end{align}
Here, ${}_2F_1(a,b,c;x)$ is the Gauss hypergeometric function.

\subsection{Collision flux}
\label{app:FP_main:col}
The term $F_\mathrm{col}$ in equation~(\ref{eq:FP_main}) describes losses of planetesimals because of physical collisions with other objects. We consider collisions of planetesimals with late-type stars and with other planetesimals of the same size. The collision cross section including gravitational focusing is $\sigma_{\mathrm{col};j} = \pi R_j^2(1 + v_{\mathrm{esc};j}^2/v_\mathrm{rel}^2)$, where $R_j\approx R_\mathrm{LT}$ in the case of collisions with late-type stars and $R_j=2R_\mathrm{a}$ in the case of collisions with other planetesimals; we adopt $R_\mathrm{a} = 10\,\mathrm{km}$ (cf. Section~\ref{sect:models:disc_models}). The escape speed $v_{\mathrm{esc};j} = (2Gm_j/R_j)^{1/2}$, where we assume $m_j=m_\mathrm{LT}$ and $m_j=2\times10^{-15}\,\mathrm{M}_\odot$ for collisions with late-type stars and planetesimals, respectively, and $v_\mathrm{rel} = ||\mathbf{v}-\mathbf{v}'||$ is the relative speed at infinity. The collision rate $\nu_{\mathrm{col};j}$ is subsequently found by replacing $n$ in $n_j \sigma_{\mathrm{col};j} v_\mathrm{rel}$ with the impactor distribution function $f_j(v')$ and integrating over all impactor velocities $v'$ (e.g. \citealt{ck78}),
\begin{align}
\nonumber \nu_{\mathrm{col};j}(v) &= \int \mathrm{d}^3v' \, f_j(v') \sigma_{\mathrm{col};j}(||\mathbf{v}-\mathbf{v}'||) ||\mathbf{v}-\mathbf{v}'|| \\
&= 2 \pi R_j^2 g_j(v) + 2 \pi G m_j R_j h_j(v).
\label{eq:col_main}
\end{align}
Here, $g_j(v)$ and $h_j(v)$ are the `Rosenbluth potentials' $h_j(v) = \int \mathrm{d}^3 v' \, f_j(v') ||\mathbf{v}-\mathbf{v}'||$ and $h_j(v) = 2\int \mathrm{d}^3 v' \, f_j(v') ||\mathbf{v}-\mathbf{v}'||^{-1}$\citep{rmj57}. We subsequently average $\nu_{\mathrm{col};j}(v)$ over a planetesimal orbit using an equation similar to equation~(\ref{eq:f_strip_enc_orb_av}) with $v=\sqrt{2[\psi(r)-\Ec]}$ and obtain $\nu_{\mathrm{col};t;j}(\Ec) = 2 \pi R_j^2 g_{t;j}(\Ec) + 2\pi G m_j R_j h_{t;j}(\Ec)$, where
\begin{align}
\nonumber g_{t;j}(\Ec) &= \frac{4\pi}{3}  \frac{4}{p(\Ec)} \int_0^{\psi^{-1}(\Ec)} \mathrm{d} r \, r^2 v(r,\Ec) \int_\Ec^{\psi(r)} \mathrm{d} \Ec' \, f_j(\Ec') \left [ \left ( \frac{\psi(r) - \Ec'}{\psi(r) - \Ec} \right )^{1/2} \left [8\psi(r) - 6 \Ec - 2 \Ec' \right ] + 8\psi(r) -2\Ec - 6\Ec' \right ]; \\
\nonumber h_{t;j}(\Ec) &= 8\pi \frac{4}{p(\Ec)} \int_0^{\psi^{-1}(\Ec)} \mathrm{d} r \, r^2 v(r,\Ec) \int_\Ec^{\psi(r)} \mathrm{d} \Ec' \, f_j(\Ec') \left ( \frac{\psi(r) - \Ec'}{\psi(r) - \Ec} \right )^{1/2} + 8\pi \int_{-\infty}^\Ec \mathrm{d} \Ec' \, f_j(\Ec')
\end{align}
(note that in the orbit-averaged expressions for the collision rate in Eq. (43) of \citealt{ck78} the stellar potential was neglected). 

The collision rate $\nu_{\mathrm{col};t;j}(\Ec)$ gives the rate of collisions with impactor $j$ for a single planetesimal at energy $\Ec$; the flux for all planetesimals at energy $\Ec$, $F_{\mathrm{col};j}$, is given by $F_{\mathrm{col};j}(\Ec,t) = \nu_{\mathrm{col};t;j}(\Ec) N_\mathrm{a}(\Ec,t)$. The combined collision flux for both types of collisions is $F_\mathrm{col}(\Ec,t) = \sum_j F_{\mathrm{col};j}(\Ec,t)$.  

In the case of planetesimal-planetesimal collisions there is a non-linear dependence of $F_{\mathrm{col};\mathrm{a}}(\Ec,t)$ on the planetesimal distribution function $f_\mathrm{a}$. To simplify the computations we use the time-independent (and therefore strictly incorrect) distribution function $f_\mathrm{a}(\Ec) = N_\mathrm{a/\star} \times f_\mathrm{LT}(\Ec)$ to compute $\nu_{\mathrm{col};t;\mathrm{a}}$ instead of the actual $f_\mathrm{a}(\Ec,t)$, the function for which equation~(\ref{eq:FP_main}) is to be solved. The former distribution function is appropriate for complete stripping of all planetesimals from the late-type stars, and without further change of the distribution function. This gives an upper limit for the importance of planetesimal-planetesimal collisions. It turns out that planetesimal-planetesimal collisions, even assuming this upper limit, are negligible compared to planetesimal-late-type star collisions; e.g., the rates for planetesimal-planetesimal collisions are unaffected compared to the main model, model 1, whereas this is not the case for late-type star-planetesimal collisions, cf. Table \ref{table:dis_rates}. This justifies our simplified treatment of planetesimal-planetesimal collisions. 

\subsubsection{Flux into the loss cone from resonant relaxation}
\label{app:FP_main:RR}
Close to the SBH the stellar potential is not perfectly spherically symmetric because the number of stars is finite. This asymmetry gives rise to resonant relaxation (RR) and results in periodic changes of the orbital angular momenta on time-scales that can be much shorter than NRR time-scale if sufficiently close to the SBH \citep{rt96}. The process of RR can potentially increase the rate of captures by the SBH arising from NRR in angular-momentum space alone (cf. Section~\ref{app:FP_main:lc}). An approximate method to model (incoherent) RR arising from a given distribution of stars in equation~(\ref{eq:FP_main}) is to include the sink term $F_\mathrm{RR}(\Ec,t)$, which is essentially the number of planetesimals with energies between $\Ec$ and $\Ec+\mathrm{d}\Ec$, divided by the time-scale for RR to decrease the angular momentum to the value corresponding to disruption by the SBH \citep{ha06,mamw11},
\begin{align}
\nonumber F_\mathrm{RR}(\Ec,t) &= \chi_\mathrm{RR} \frac{N_\mathrm{a}(\Ec,t)}{\left | \ln\left[\Rc_\mathrm{lc}(\Ec)^{1/2}\right] \right | \,T_\mathrm{RR}}; \\
T_\mathrm{RR} &= \beta_s^{-2} \left (\frac{M_\bullet}{m_\star} \right )^2 N(\Ec)^{-1} \frac{P(\Ec)^2}{t_\mathrm{coh}(\Ec)}.
\label{eq:F_RR}
\end{align}
Here, $\beta_s$ describes the efficiency of RR; numerical investigations indicate that $\beta_s \approx 1.6 e$ in the Newtonian regime \citep{gh07,eka09,hpm14}. Averaging this quantity over a thermal distribution yields $\beta_s \approx 1.1$; here we adopt $\beta_s=1$. The factor $|\ln[\Rc^{1/2}_\mathrm{lc}(\Ec)]|$ is approximately the typical number of relaxation time-scales for the angular momentum to decrease from the maximal value corresponding to a circular orbit, to the value corresponding to capture by the SBH. 

The quantity $t_\mathrm{coh}(\Ec)$ in equation~(\ref{eq:F_RR}) is the coherence time-scale, i.e. the typical time-scale for field stars to change their orientation with respect to a test orbit with energy $\Ec$. Here, we adopt $t^{-1}_\mathrm{coh}(\Ec) = t_\mathrm{MP}^{-1}(\Ec) + t_\mathrm{GR}^{-1}(\Ec)$, where $t_\mathrm{MP}(\Ec) = [M_\bullet/M_\star(\Ec)]P(\Ec)$ is an estimate of the (angular-momentum-averaged) Newtonian mass precession time-scale and $t_\mathrm{GR}(\Ec) = (1/24) (c^2/\Ec) P(\Ec)$ is the (angular-momentum-averaged) relativistic precession time-scale, assuming a thermal eccentricity distribution. 

At low angular momenta the efficiency of RR is strongly reduced because in-plane relativistic precession tends to reduce the efficiency of the torques arising from the $\sqrt{N}$-asymmetry. This effect, known as the Schwarzschild barrier, can reduce the flux implied by equation~(\ref{eq:F_RR}) by at least an order of magnitude \citep{mamw11}. It remains unclear, however, how the details of equation~(\ref{eq:F_RR}) are affected by the SB. To take into account this uncertainty we include in equation~(\ref{eq:F_RR}) the dimensionless (and poorly-constrained) ad hoc factor $\chi$, and adopt two values of $\chi_\mathrm{RR}$: $\chi_\mathrm{RR}=0.1$ and $\chi_\mathrm{RR}=1$.

\section{Approximate semi-analytic solutions to the time-dependent Fokker-Planck equation}
\label{app:FP_semi_an}
We solve equation~(\ref{eq:FP_main}) with $F_\mathrm{col} = F_\mathrm{RR} = 0$ and assume that a steady-state is present in energy at all times, i.e. $f_\mathrm{a}(\Ec,t) = f_0 g(t)$, where $f_0$ is a constant and $g(t)$ a time-dependent function with $g(0)=1$ (cf. Section~\ref{sect:dyn_ff_ast:results:semi_an}). In this case $\partial \mathcal{F}_\Ec/\partial \Ec=0$; integrating both sides of equation~(\ref{eq:FP_main}) with respect to energy, the latter equation can be written as
\begin{align}
C_1 \frac{\partial g}{\partial t} = N_\mathrm{a/\star} \frac{\partial h}{\partial t} - C_2 g(t),
\label{eq:FP_semi_an_1}
\end{align}
with
\begin{align}
\nonumber C_1 \equiv f_0 4 \pi^2 \int \mathrm{d} \Ec \, p(\Ec); \quad C_2 \equiv C_1 \frac{\int \mathrm{d} \Ec \, S_\mathrm{lc}}{4\pi^2 \int \mathrm{d} \Ec \, p(\Ec)}; \quad h(t) \equiv \int \mathrm{d} \Ec \, N_\mathrm{LT}(\Ec) f_\mathrm{strip}(\Ec,t),
\end{align}
and $S_\mathrm{lc}(\Ec)$ is defined via $F_\mathrm{lc}(\Ec,t) = S_\mathrm{lc}(\Ec) f_\mathrm{a}(\Ec,t)$ (cf. equation~\ref{eq:F_lc}). In the case of formation in a cloud $C_1$, and thereby $f_0$, can be estimated from $C_1\sim N_\mathrm{a/\star} \int \mathrm{d} \Ec N_\mathrm{LT}(\Ec)$; in the case of formation in discs, $C_1 \sim N_\mathrm{a/\star} \int \mathrm{d} \Ec N_\mathrm{LT}(\Ec) f_\mathrm{strip;SBH}(\Ec)$. Equation~(\ref{eq:FP_semi_an_1}) can be solved for $g(t)$ by separation of variables; writing $g(t)=g_H(t)g_I(t)$, the homogenous solution is
\begin{align}
g_H(t) = g_H(0) \exp \left ( \frac{-t}{\tau} \right ),
\end{align}
where $\tau \equiv C_1/C_2$. Substituting the latter solution into equation~(\ref{eq:FP_semi_an_1}) and solving for $g_I(t)$, we find
\begin{align}
g_I(t) = \frac{N_\mathrm{a/\star}}{C_1 g_H(0)} \int_0^t \mathrm{d} t' \, \frac{\partial h}{\partial t'} \exp \left ( \frac{t'}{\tau} \right ) + C,
\end{align}
where $C$ is an integration constant. Setting $g(0) = 1$, the complete solution is
\begin{align}
g(t) = \exp \left ( \frac{-t}{\tau} \right ) \left [ 1 + \frac{N_\mathrm{a/\star}}{C_1} \int_0^t \mathrm{d} t' \, \frac{\partial h}{\partial t'} \exp \left ( \frac{t'}{\tau} \right ) \right ].
\label{eq:FP_semi_an_g}
\end{align}
The implied disruption flux is $F_\mathrm{dis}(t) \sim f_0 g(t) \int \mathrm{d} \Ec \, S_\mathrm{lc}(\Ec)$ (cf. equation~\ref{eq:F_lc}) and is shown in Fig. \ref{fig:F_lc_t_semi_an}, where in case formation in a  cloud we set $h(t)=0$. 

\section{Scaling of the disruption rate}
\label{app:scaling}
To obtain the scaling of the disruption rate with $N_\mathrm{a/\star}$, $m_\star$, $M_\bullet$ and $N_0$ (cf. Section~\ref{discussion:scaling}) we adopt the approximate semi-analytic solution to equation~(\ref{eq:FP_main}) as described in Appendix \ref{app:FP_semi_an}. We approximate the effect of stripping by gravitational encounters as an instantaneous process at all energies, i.e.
\begin{align}
h(t) \sim \left \{ \begin{array}{ll}
0, & t<t_c; \\
\int \mathrm{d} \Ec \, N_\star(\Ec), & t \geq t_c, \\
\end{array} \right.
\end{align}
where $N_\star(\Ec)$ is the number of stars with energies between $\Ec$ and $\Ec+\mathrm{d}\Ec$ and $t_c$ is the stripping time-scale; in the GC, $t_c\sim 100 \, \mathrm{Myr}$ (cf. Section~\ref{sect:stripping:enc}). Integrating by parts, we subsequently obtain
\begin{align}
\int_0^t \mathrm{d}t' \frac{\partial h}{\partial t'} \exp \left ( \frac{t'}{\tau} \right ) \sim \left \{ \begin{array}{ll}
0, & t < t_c; \\
\exp \left ( t_c/\tau \right ) \int \mathrm{d} \Ec \, N_\star(\Ec), & t \geq t_c. \\
\end{array} \right. 
\end{align}
Neglecting the initial stripping by the SBH, $C_1 \sim N_\mathrm{a/\star} \int \mathrm{d} \Ec N_\star(\Ec)$ (cf. Appendix \ref{app:FP_semi_an}), and equation~(\ref{eq:FP_semi_an_g}) yields, for $t\geq t_c$,
\begin{align}
g(t) \sim \exp \left ( \frac{-t}{\tau} \right ) \left [ 1 + \exp \left ( \frac{t_c}{\tau} \right ) \right ].
\label{eq:scaling_der_g}
\end{align}
Note that $\tau \equiv C_1/C_2 \equiv 4 \pi^2 \int \mathrm{d} \Ec \, p(\Ec) / \int \mathrm{d} \Ec S_\mathrm{lc}(\Ec)$ is independent of $N_\mathrm{a/\star}$, therefore $g(t)$ in equation~(\ref{eq:scaling_der_g}) is also independent of $N_\mathrm{a/\star}$. From $f_\mathrm{a}(\Ec,t) = f_0 g(t)$ and $f_0\propto N_\mathrm{a/\star}$ it subsequently follows that the disruption flux (cf. equation~\ref{eq:F_lc}) scales linearly with $N_\mathrm{a/\star}$, as might be intuitively expected.

We assume a power-law stellar number density distribution, $n_\star=n_0(r/r_0)^{-\gamma}$, where $n_0 = N_0 (3-\gamma)/(4\pi r_0^3)$. The distribution function is then given by $f_\star(\Ec) = f_0 \Ec^{\gamma-3/2}$, where $f_0$ is related to $n_0$ via an equation similar to equation~(\ref{eq:f0_n0}). Neglecting, for simplicity, the integrals $\bar{I}_{1/2}(\Ec)$ and $\bar{I}_{3/2}(\Ec)$ in equation~(\ref{eq:mu_e}), and setting the integration limits in the energy integral to $\Ec_h < \Ec < \infty$, where $\Ec_h = [GM_\bullet/r_0] [ (2M_\bullet/m_\star) (1/N_0) ]^{1/(\gamma-3)}$ is the energy at the radius of influence neglecting the stellar potential, the energy integral of $S_\mathrm{lc}$ (cf. equation~\ref{eq:F_lc}) is approximately given by
\begin{align}
\int_{\Ec_h}^{\infty} \mathrm{d} \Ec \, S_\mathrm{lc}(\Ec) \sim 4 \pi^2 \frac{32}{3} \pi^2 \frac{5\pi}{16} \frac{1}{\sqrt{2}} \frac{1}{3-\gamma} (Gm_\star)^2 (G M_\bullet)^3 \ln (\Lambda) \Ec_h^{\gamma-3} f_0,
\end{align}
where we neglected the factor $\log( R_0^{-1})^{-1}$, the integrals $\bar{I}_{n/2}(\Ec)$ and assumed $\gamma\leq 2$. With neglect of the stellar potential the function $p(\Ec)$ is given by $p(\Ec) = (\sqrt{2} \pi/4)(GM_\bullet)^3 \Ec^{-5/2}$ (e.g. \citealt[][eq. 5.182a]{bookmerritt13}). The time-scale $\tau$ is therefore approximately given by
\begin{align}
\tau^{-1} \sim \frac{5}{4\sqrt{2\pi}} \frac{\Gamma(\gamma+1)}{\Gamma \left (\gamma-\frac{1}{2} \right )} \log(\Lambda) (Gm_\star)^2 \left ( GM_\bullet r_0 \right )^{-3/2} \left ( \frac{2M_\bullet}{m_\star} \frac{1}{N_0} \right )^\frac{\gamma-3/2}{\gamma-3} N_0.
\end{align}

Setting $\int_{\Ec_h}^{\infty} \mathrm{d} \Ec \, N_\star(\Ec) \sim N_0$, we find for the disruption flux
\begin{align}
F_\mathrm{dis}(t) \sim f_0 g(t) \int \mathrm{d} \Ec \, S_\mathrm{lc}(\Ec) \sim \frac{C_1 g(t)}{\tau} \sim \frac{N_\mathrm{a/\star} N_0}{\tau} \exp(-t/\tau) \left [ 1 + \exp(t_c/\tau) \right ].
\end{align}

\end{document}